\documentclass[preprint,aps,12pt,notitlepage,nofootinbib,tightenlines]{revtex4}
\usepackage{setspace}
\usepackage{placeins}
\usepackage{amsmath}
\usepackage{caption}
\usepackage{subcaption}
\usepackage[per-mode=symbol]{siunitx}
\usepackage{booktabs}
\usepackage{multirow}
\usepackage{graphicx}
\usepackage{enumitem}
\usepackage{underscore}
\usepackage{bm}
\usepackage{times}
\usepackage{braket}
\usepackage{color}
\usepackage{epsfig}
\usepackage{slashed}
\usepackage{hyperref}
\usepackage{array}
\usepackage{float}
\newcommand{\beq}{\begin{eqnarray}}
\newcommand{\eeq}{\end{eqnarray}}
\newcommand{\be}{\begin{equation}\begin{aligned}}
\newcommand{\ee}{\end{aligned}\end{equation}}

\definecolor{Red}{rgb}{1.,0.,0.}

\definecolor{Blue}{rgb}{0.,0.,1.}

\definecolor{nicered}{rgb}{0.7,0.1,0.1}
\definecolor{nicegreen}{rgb}{0.1,0.5,0.1}
\def\lsim{ {\ \lower-1.2pt\vbox{\hbox{\rlap{$<$}\lower6pt\vbox{\hbox{$\sim$}}}}\ } }
\def\gsim{ {\ \lower-1.2pt\vbox{\hbox{\rlap{$>$}\lower6pt\vbox{\hbox{$\sim$}}}}\ } }
\hypersetup{colorlinks,citecolor=nicegreen,linkcolor=nicered}

\begin{document}
\title{Probing first-generation doublet vector-like leptons via single production at the 3 TeV CLIC}
\author{Yin-Hao Gao$^{1}$, Yao-Bei Liu$^{1,2}$\footnote{E-mail: liuyaobei@hist.edu.cn}, Stefano Moretti$^{3,4}$\footnote{E-mail: stefano.moretti@cern.ch} }
\affiliation{1. Henan Institute of Science and Technology, Xinxiang 453003, China\\
2. School of Electro-Mechanical Engineering, Zhongyuan University of Science and Technology, Xuchang 461000, China\\
3. School of Physics and Astronomy, University of Southampton, Highfield, Southampton SO17 1BJ, United Kingdom \\
4. Department of Physics and Astronomy, Uppsala University, Box 516, 751 20 Uppsala, Sweden}
\vspace*{0.5truecm}
\begin{abstract}
{\small \begin{singlespace}
The single production of first-generation $\mathrm{SU}(2)_L$ doublet
vector-like leptons (VLLs) at the Compact Linear Collider (CLIC) with
$\sqrt{s}=3~\mathrm{TeV}$ is studied. The VLL doublet consists of a
neutral state $N$ and a charged state $E^\pm$, whose single-production
processes are investigated through
$e^+e^-\rightarrow \nu_e\bar N+\bar\nu_e N$ and
$e^+e^-\rightarrow e^\pm E^\mp$. The neutral VLL is probed via the
decay channel $N\rightarrow W^-e^+$
($\bar N\rightarrow W^+e^-$), followed by $W^\pm\rightarrow jj$,
resulting in the $e^\pm+2j+\slashed{E}_T$ final state. The charged VLL
is explored through the decay modes $E^\pm\rightarrow e^\pm h$ with
$h\rightarrow b\bar b$ and $E^\pm\rightarrow e^\pm Z$ with
$Z\rightarrow q\bar q$, yielding the $e^+e^-+2b$ and
$e^+e^-+2j$ signatures, respectively. Signal and background events are
simulated by incorporating initial-state radiation, beamstrahlung, and
detector effects whereas  cut-based analyses are performed for integrated
luminosities of $1$, $3$, and $5~\mathrm{ab}^{-1}$. The expected
$2\sigma$ exclusion and $5\sigma$ discovery sensitivities are obtained
in the VLL mass--mixing parameter space. Among the investigated
channels, the neutral-VLL production mode exhibits the highest
sensitivity. For an integrated luminosity of $1~\mathrm{ab}^{-1}$, the
$2\sigma$ exclusion sensitivity reaches $\varepsilon\simeq0.013$ at
$M_N=1.3~\mathrm{TeV}$ and $\varepsilon\simeq0.065$ at
$M_N=2.9~\mathrm{TeV}$, while the corresponding $5\sigma$ discovery
sensitivity extends from approximately $\varepsilon\simeq0.02$ to
$0.10$. The charged-VLL channels provide complementary probes, with
$5\sigma$ discovery sensitivities reaching VLL masses of about
$2.9~\mathrm{TeV}$ and $2.8~\mathrm{TeV}$ for the
$E^\pm\rightarrow e^\pm h$ and $E^\pm\rightarrow e^\pm Z$ channels,
respectively, at $\mathcal{L}=5~\mathrm{ab}^{-1}$ within the parameter
space consistent with electroweak precision constraints. These results
demonstrate that a 3 TeV CLIC can provide sensitive probes of
first-generation doublet VLLs over a wide mass range, including
scenarios with small VLL--electron mixing.
\end{singlespace}}
\end{abstract}

\maketitle
\newpage

\section{Introduction}

Vector-like leptons (VLLs) are among the well-motivated extensions of
the Standard Model (SM) and have been extensively explored in various
new physics scenarios, including extra-dimensional setups
~\cite{Huang:2012kz,Kong:2010qd}, supersymmetric theories
~\cite{Graham:2009gy,Endo:2011xq,Martin:2012dg,Endo:2012cc,
Fischler:2013tva}, composite Higgs models
~\cite{DeCurtis:2018iqd,DeCurtis:2018zvh}, and other extensions of the SM
~\cite{He:1999vp,Wang:2013jwa,He:2001fz,He:2014ora}. Their vector-like
nature allows them to arise naturally in theories addressing several
fundamental issues in particle physics, such as the hierarchy problem
associated with the Higgs sector~\cite{Arkani-Hamed:2012dcq}, dark
matter phenomenology~\cite{Schwaller:2013hqa,Halverson:2014nwa,
Bahrami:2016has,Bhattacharya:2018fus}, electroweak (EW) vacuum stability
~\cite{Xiao:2014kba,Cingiloglu:2024vdh}, and precision observables,
including the electron and muon anomalous magnetic moments
~\cite{Hiller:2019mou,DeJesus:2020yqx,Frank:2020smf,Dermisek:2021ajd,
Brune:2022rlo,Guedes:2022cfy,Erdelyi:2025axy,Athron:2025ets} and the
reported deviation in the $W$-boson mass measurement
~\cite{He:2022zjz,Kawamura:2022fhm}. These theoretical motivations,
together with their distinctive collider signatures, make VLLs an
important target in searches for physics beyond the SM.

Unlike the chiral fermions in the SM, VLLs contain left- and
right-handed components with identical transformation properties under
the SM gauge symmetry. Consequently, they can acquire gauge-invariant
Dirac mass terms independent of EW symmetry breaking (EWSB), leading
to a characteristic mass generation mechanism distinct from that of SM
fermions. This feature alleviates some constraints from EW
precision observables (EWPOs)~\cite{delAguila:2008pw,Ishiwata:2013gma} and Higgs
coupling measurements~\cite{Kearney:2012zi,Altmannshofer:2013zba,
Falkowski:2013jya}, while their mixing with SM leptons gives rise to
rich production and decay signatures at high-energy colliders.
Extensive phenomenological studies have investigated the collider
implications of VLLs at both current and future facilities (see, e.g.,
Refs.~\cite{Ellis:2014dza,Ishiwata:2015cga,Dermisek:2014cia,
Crivellin:2020ebi,Endo:2020tkb,Chakrabarty:2020jro,Guedes:2021oqx,
Raju:2022zlv,Li:2023mrw,Dermisek:2014qca,Dermisek:2015oja,
Kumar:2015tna,Bhattiprolu:2019vdu,Dermisek:2015hue,Chen:2016lsr,
Kawamura:2019rth,Freitas:2020ttd,OsmanAcar:2021plv,Kawamura:2021ygg,
Bonilla:2021ize,Baspehlivan:2022qet,Cao:2023smj,Bernreuther:2023uxh,
Kawamura:2023zuo,Bigaran:2023ris,Bissmann:2020lge,Morais:2021ead}).
Depending on the mixing pattern, VLLs are commonly classified as vector-like electrons, muons, or tauons, corresponding to their mixing with the first-, second-, or third-generation SM leptons ($\ell=e,\mu,\tau$). In this work, we focus on the first-generation doublet VLL scenario, where both the charged component $E^\pm$ and the neutral component $N$
are assumed to mix with the electron sector. Current collider searches have mainly concentrated on pair production of VLLs at the LHC, particularly in final states involving third-generation leptons and EW gauge bosons~\cite{ATLAS:2015qoy,ATLAS:2023sbu,CMS:2019hsm,CMS:2022nty,CMS:2022cpe,CMS:2024bni}. Recent ATLAS searches have established stringent 95\% confidence-level (CL) lower bounds, excluding doublet vector-like electrons and muons below approximately $1220$~GeV and $1270$~GeV, respectively, while the limits on singlet VLL scenarios are considerably weaker~\cite{ATLAS:2024mrr}.

Future high-energy $e^+e^-$ colliders, such as the International
Linear Collider (ILC)~\cite{ILC:2013jhg,ILCInternationalDevelopmentTeam:2022izu}
and the Compact Linear Collider (CLIC)~\cite{Linssen:2012hp,CLICDetector:2013tfe,
Franceschini:2019zsg}, offer a complementary approach to searching for VLLs.
Compared with hadron colliders, lepton colliders provide a cleaner
experimental environment, a well-defined initial state, and enhanced
control over EW processes. The phenomenology of VLLs at future
lepton colliders has been extensively studied for various EW
representations and flavor structures~\cite{Das:2020gnt,Das:2020uer,
Yang:2021dtc,Shang:2021mgn,Bhattacharya:2021ltd,Shang:2023rfv,
Bhattiprolu:2023yxa,Yue:2024sds,Yue:2024ftz,Yue:2025xvk,Shen:2025mxe,
Li:2025epjc,Yue:2026tbc,Sagheer:2026eql}. In addition, the single
production of singlet VLLs with third-generation mixing has been explored
at the LHC using machine-learning approaches~\cite{Cui:2026wwo}.
Recently, we investigated the single production of first-generation
singlet charged VLLs at $e^+e^-$ colliders with
$\sqrt{s}=1$ and $1.5$~TeV~\cite{Liu:2026eyn}. However, current LHC
searches exclude doublet vector-like electrons below approximately
$1.22$~TeV~\cite{ATLAS:2024mrr}, leaving essentially no accessible mass
range at $\sqrt{s}=1$~TeV and only a limited one at
$\sqrt{s}=1.5$~TeV. We therefore focus here on the $3$~TeV stage of CLIC,
where single production can probe VLL masses well above the
pair-production threshold $M_{\rm VLL}\lesssim\sqrt{s}/2$.

Unlike the singlet case, the doublet representation contains both a
charged state $E^\pm$ and a neutral state $N$, giving rise to
complementary search channels. Accordingly, we investigate the single
production of first-generation doublet VLLs at the $3$~TeV CLIC through
three complementary channels:
\begin{enumerate}[label=(\roman*)]
\item
$e^+e^- \to \nu_e\bar{N}+\bar{\nu}_e N$,
followed by
$N\to W^-e^+$ and $\bar{N}\to W^+e^-$,
with $W^\pm\to q\bar{q}'$, yielding the
$e^\pm+2j+\slashed{E}_T$ final state;

\item
$e^+e^- \to e^\pm E^\mp$,
followed by
$E^\pm\to e^\pm h$,
with $h\to b\bar{b}$, yielding the
$e^+e^-+2b$ final state;

\item
$e^+e^- \to e^\pm E^\mp$,
followed by
$E^\pm\to e^\pm Z$,
with $Z\to q\bar{q}$, yielding the
$e^+e^-+2j$ final state.
\end{enumerate}
The effects of initial-state radiation (ISR) and beamstrahlung are
included, and a Monte Carlo (MC)  analysis is performed at
$\sqrt{s}=3~\mathrm{TeV}$ for integrated luminosities of
$\mathcal{L}=1$, $3$, and $5~\mathrm{ab}^{-1}$. The resulting exclusion
and discovery sensitivities are presented in the VLL mass--mixing
parameter plane.

The remainder of this paper is organized as follows. In
Sec.~\ref{sec:model}, we introduce the theoretical framework and
summarize the production  and decay properties of the doublet VLLs.
In Sec.~\ref{sec:analysis}, we describe the single-production processes
at CLIC, the MC simulation, the event selection strategy, and the
resulting exclusion and discovery sensitivities. Finally, we present
our conclusions in Sec.~\ref{sec:conclusion}.

\section{Theoretical framework}
\label{sec:model}

We consider the $\mathrm{SU}(2)_L$ doublet VLL model
introduced in Ref.~\cite{Kumar:2015tna}, with the VLLs mixing only with
the first-generation SM leptons. The new vector-like doublet is given by
\begin{eqnarray}
\label{eq:2.1}
L'=
\begin{pmatrix}
N\\
E^-
\end{pmatrix},
\end{eqnarray}
where both chiral components transform as
$L'_{L,R}\sim(\mathbf{1},\mathbf{2},-\tfrac{1}{2})$
under the SM gauge group.

The interaction between the VLL doublet and the first-generation SM
leptons is generated through Yukawa couplings with the Higgs doublet.
The relevant terms in the Lagrangian are
\begin{eqnarray}
\label{eq:2.2}
-\mathcal{L}
&=&
M_L\bar L'_L L'_R
+\varepsilon\,\bar L'_L H e_R
+y_e\,\bar L_e H e_R
+\mathrm{H.c.},
\end{eqnarray}
where $L_e=(\nu_e,e)_L$ denotes the SM lepton doublet, $y_e$ is the SM
electron Yukawa coupling, and $\varepsilon$ parametrizes the
VLL--electron mixing strength.

After EWSB with
$\langle H\rangle=v\simeq174$ GeV, the charged-lepton mass matrix in the
$(e,E^-)$ basis is
\begin{equation}
\mathcal{M}_{eE}
=
\begin{pmatrix}
y_ev&0\\
\varepsilon v&M_L
\end{pmatrix}.
\end{equation}
The physical mass eigenstates are obtained by diagonalizing this matrix.
We denote the VLL--electron mixing angle by $\theta_\ell$, where the
subscript $\ell$ refers to the lepton sector. In the small-mixing limit,
$\varepsilon v\ll M_L$, it satisfies
\begin{equation}
\sin\theta_\ell\simeq\frac{\varepsilon v}{M_L}.
\end{equation}
Throughout this work, we focus on this perturbative mixing regime.

At tree level and in the absence of mixing, the charged and neutral
components of the VLL doublet are degenerate,
$M_N=M_E=M_L$. After including the mixing effect, the charged VLL mass
becomes
$M_E\simeq\sqrt{M_L^2+\varepsilon^2v^2}$,
while the neutral component remains unchanged,
$M_N=M_L$. The resulting mass splitting,
\begin{equation}
M_E-M_N\simeq\frac{\varepsilon^2v^2}{2M_L},
\end{equation}
is negligible for TeV-scale VLL masses considered here. Electroweak
radiative corrections induce an additional splitting of order a few
hundred MeV~\cite{Thomas:1998wy}.

To leading order in the mixing parameter $\varepsilon$, the relevant
interactions between the VLLs and the SM gauge and Higgs bosons are
given by~\cite{Kumar:2015tna}
\begin{eqnarray}
\mathcal{L}_{\mathrm{int}}
=&
g_{\bar eN}^{W^+}
\left[
W_\mu^+
(\bar e^\dagger\bar\sigma^\mu N)
+
W_\mu^-
(\bar N^\dagger\bar\sigma^\mu e)
\right]
\nonumber\\
&+
g_{\bar eE}^{Z}
Z_\mu
(\bar e^\dagger\bar\sigma^\mu E
+
\bar E^\dagger\bar\sigma^\mu e)
+
(y_{E\bar e}^{h}hE\bar e+\mathrm{H.c.}),
\end{eqnarray}
with the effective couplings
\begin{eqnarray}
g_{\bar eN}^{W^+}
&=&
-\varepsilon\frac{M_W}{M_L},
\nonumber\\
g_{\bar eE}^{Z}
&=&
-\varepsilon\frac{M_Z}{\sqrt2 M_L},
\nonumber\\
y_{E\bar e}^{h}
&=&
-\frac{\varepsilon}{\sqrt2}.
\end{eqnarray}

In this doublet VLL scenario, the decay mode
$E^\pm\rightarrow W^\pm\nu_e$ is absent at leading order. Therefore,
the charged VLL decays through the channels
$E^\pm\rightarrow e^\pm Z$ and $E^\pm\rightarrow e^\pm h$, while the
neutral component decays through
$N\rightarrow W^-e^+$ and $\bar N\rightarrow W^+e^-$.

The corresponding decay widths are
\begin{eqnarray}
\Gamma(N\rightarrow W^-e^+)
&=&
\frac{M_N}{32\pi}(1-r_W)^2(2+1/r_W)
|g_{\bar eN}^{W^+}|^2,
\nonumber\\
\Gamma(E^-\rightarrow e^-Z)
&=&
\frac{M_E}{32\pi}(1-r_Z)^2(2+1/r_Z)
|g_{\bar eE}^{Z}|^2,
\nonumber\\
\Gamma(E^-\rightarrow e^-h)
&=&
\frac{M_E}{32\pi}(1-r_h)^2
|y_{\bar eE}^{h}|^2,
\end{eqnarray}
where $r_X=M_X^2/M_N^2$ for neutral VLL decays and
$r_X=M_X^2/M_E^2$ for charged VLL decays.

In the heavy-mass limit, $M_L\gg M_Z,M_h$, the charged VLL decay widths
satisfy
\[
\Gamma(E^-\rightarrow e^-Z)
\simeq
\Gamma(E^-\rightarrow e^-h),
\]
as expected from the Goldstone boson equivalence theorem.

The mixing between VLLs and SM leptons modifies the EW gauge
couplings and is therefore constrained by EWPOs~\cite{Cynolter:2008ea,deBlas:2013gla,Adhikary:2024esf,deBlas:2025pco,Daberstiel:2026wxe}.
The corresponding limits depend on the EW representation of the
VLLs and the SM lepton generation involved. A recent analysis
including precision measurements of the $Z$-lepton couplings and the
$W$-boson mass finds that the mixing of an $SU(2)_L$ doublet VLL with
first-generation charged leptons is constrained at the few-percent
level~\cite{Daberstiel:2026wxe}. In this work, we adopt the bound from
Ref.~\cite{deBlas:2013gla},
$\sin\theta_\ell<0.02$, which translates into
\begin{equation}
\varepsilon < 0.02\frac{M_L}{v}.
\end{equation}

Finally, sufficiently small VLL--SM mixing can lead to long-lived VLLs
with displaced signatures. However, in the parameter region considered
in our collider analysis, the total decay width scales approximately as
$\Gamma_{\rm VLL}\sim\varepsilon^2M_L/(32\pi)$,
resulting in microscopic decay lengths for TeV-scale VLL masses. Hence,
the VLL decays can be treated as prompt.

\section{Signature of first-generation doublet VLLs at the CLIC}
\label{sec:analysis}

\subsection{Single production at the $e^+e^-$ collider}

\begin{figure}[ht]
\centering
\includegraphics[width = 14cm ]{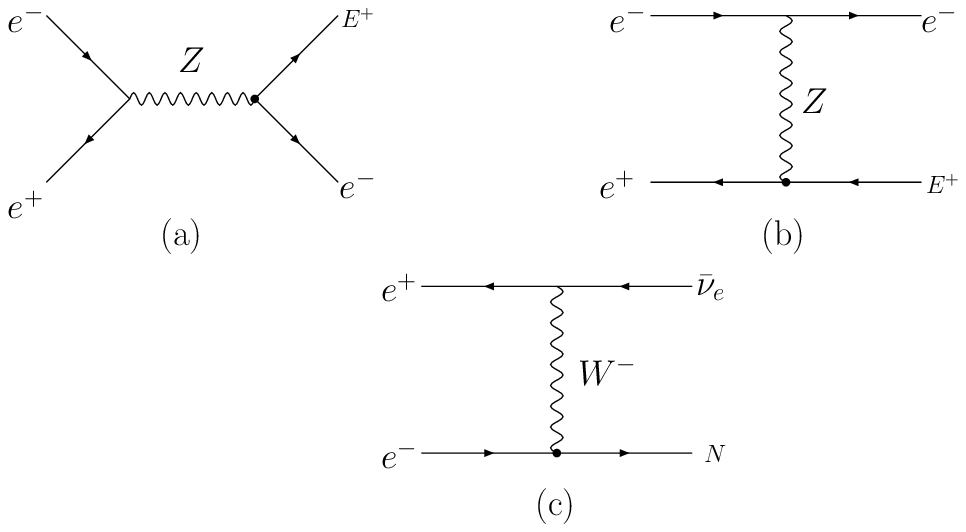}
\vspace{-12cm}
\caption{Representative Feynman diagrams for single production of first-generation
doublet VLLs. Panels (a) and (b) correspond to the charged process
$e^+e^-\to e^-E^+$, while panel (c) corresponds to the neutral process
$e^+e^-\to\bar\nu_e N$.}
\label{fig1}
\end{figure}

Representative Feynman diagrams for the single production of first-generation doublet VLLs are shown in Fig.~\ref{fig1}. Panels (a) and (b) correspond to the charged-VLL production process
$e^+e^-\to e^-E^+$,
whereas panel (c) illustrates the neutral-VLL production process
$e^+e^-\to\bar\nu_e N$.
The corresponding charge-conjugate channels are included throughout the numerical analysis.

\begin{figure}[h]
\begin{center}
\vspace{-0.5cm}
\centerline{\hspace{1.0cm}\epsfxsize=9cm\epsffile{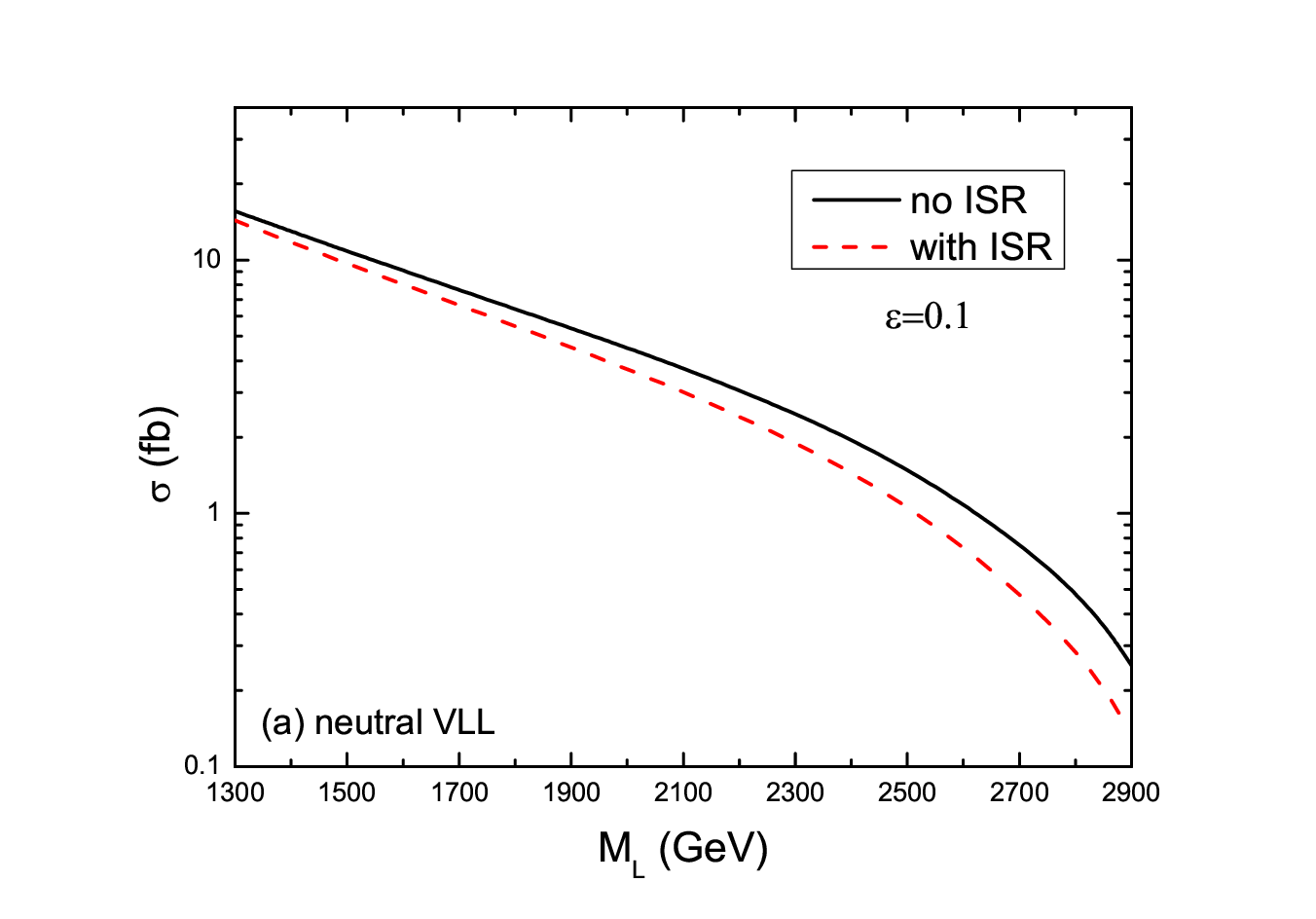}
\hspace{-1.0cm}\epsfxsize=9cm\epsffile{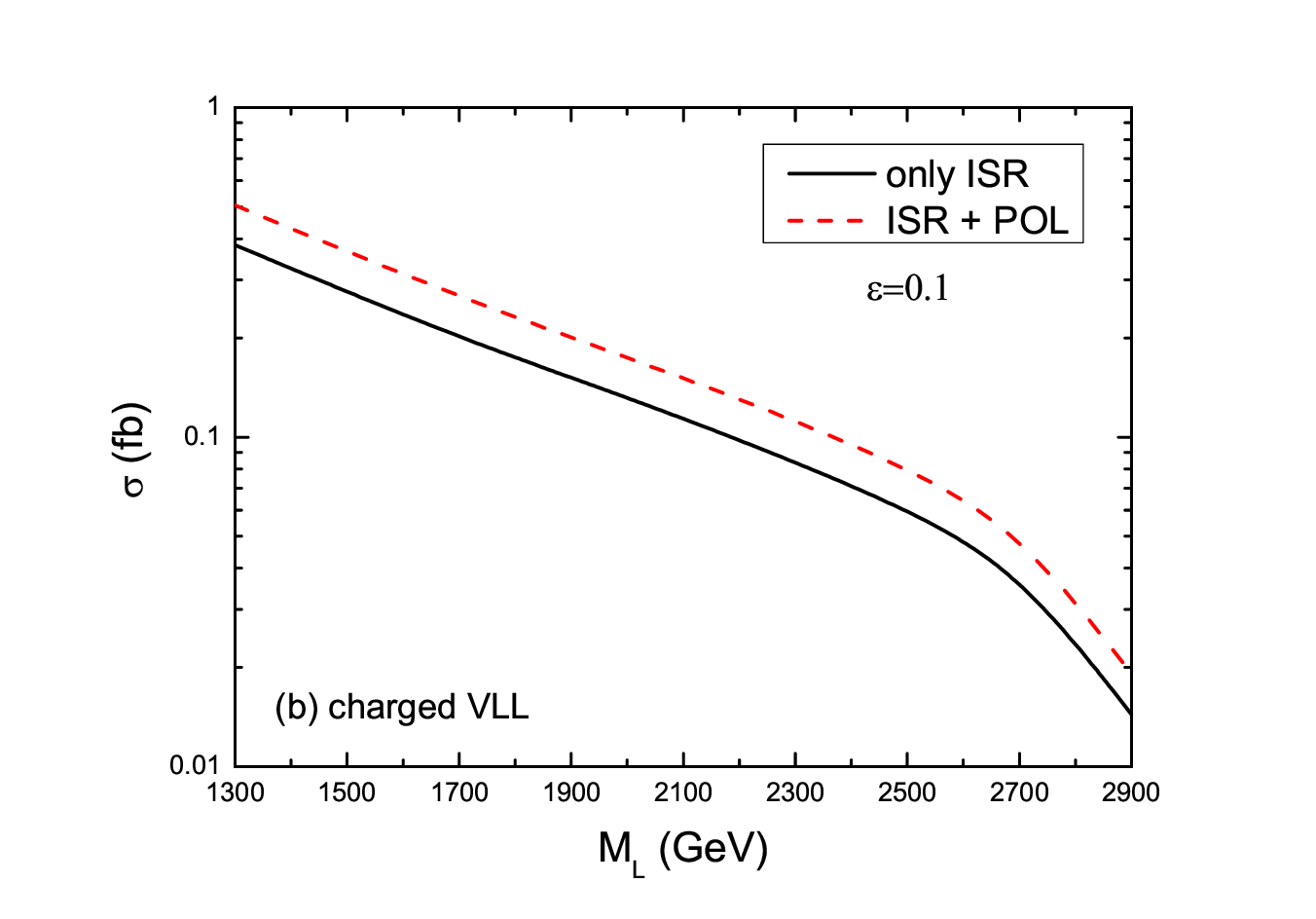}}
\caption{Production cross sections for (a) $e^+e^-\to\nu_e\bar N+\bar\nu_eN$ and (b) $e^+e^-\to e^\pm E^\mp$ as functions of the VLL mass at $\sqrt{s}=3$~TeV for $\varepsilon=0.1$, including ISR and beamstrahlung effects. For the charged channel, the polarized beam configuration $(P_{e^-},P_{e^+})=(0.8,0)$ is assumed.}
\label{fig2}
\end{center}
\end{figure}

Figure~\ref{fig2} shows the production cross sections of the neutral and charged VLL channels as functions of the VLL mass at $\sqrt{s}=3$~TeV. We take $\varepsilon=0.1$ as a representative benchmark value. This choice satisfies the EWPO constraints discussed in Sec.~\ref{sec:model} throughout the mass range considered; at the lower end, $M_L=1.3$~TeV, it corresponds to $\sin\theta_\ell\simeq0.013$, below the adopted bound $\sin\theta_\ell<0.02$. Motivated by the current LHC constraints on first-generation doublet
VLLs~\cite{ATLAS:2024mrr}, we consider the mass range
\begin{equation}
1300~{\rm GeV} \leq M_L \leq 2900~{\rm GeV}.
\end{equation}
The lower end of this interval is chosen just above the current direct
lower bound on doublet vector-like electrons, while the upper end
extends close to the kinematic limit for single production at the
$3$~TeV CLIC. This choice therefore allows us to explore essentially
the full presently allowed mass range that is kinematically accessible
at this collider energy.

The effects of ISR and beamstrahlung are included using the 3 TeV CLIC beam spectrum implemented in \texttt{MadGraph5_aMC@NLO}~\cite{Alwall:2014hca}. Since the single-production cross sections scale approximately as $\sigma_{\rm single}\propto\varepsilon^2$, the results obtained for the benchmark value $\varepsilon=0.1$ can be straightforwardly rescaled to other values of the mixing parameter.

For the neutral channel,
$e^+e^-\rightarrow\nu_e\bar N+\bar\nu_eN$,
the production cross section decreases from approximately $14.33$~fb at
$M_N=1300$~GeV to $0.13$~fb at $M_N=2900$~GeV. After including all
relevant diagrams and summing over the charge-conjugate final states, the
inclusive production rate exhibits negligible dependence on the
electron-beam polarization.

For the charged channel,
$e^+e^-\to e^\pm E^\mp$,
a beam-polarization configuration of
$(P_{e^-},P_{e^+})=(0.8,0)$
is considered. This corresponds to a predominantly right-handed electron
beam with an unpolarized positron beam and enhances the signal rate
compared with the unpolarized case. For example, at
$M_E=1300$~GeV, the cross section increases from $0.35$~fb for
unpolarized beams to $0.47$~fb for $P_{e^-}=0.8$, corresponding to an
enhancement factor of approximately $1.32$. Over the mass range considered,
the charged-VLL production cross section decreases rapidly as the
kinematic threshold is approached.

The different polarization behaviors of the neutral and charged channels
originate from their distinct helicity structures. Defining the
electron-beam polarization as
\[
P_{e^-}=\frac{N_R-N_L}{N_R+N_L},
\]
and taking the positron beam to be unpolarized, the polarized cross section
can be expressed as
\[
\sigma(P_{e^-})=
\frac{1-P_{e^-}}{2}\sigma_L+
\frac{1+P_{e^-}}{2}\sigma_R,
\]
where $\sigma_L$ and $\sigma_R$ denote the cross sections for fully
left-handed and right-handed electron beams, respectively.

For the neutral-VLL production channel, the two charge-conjugate processes
select complementary electron helicity configurations,
\[
\sigma(e^+e^-\to\bar\nu_e N)=
\frac{1+P_{e^-}}{2}\sigma_0,
\qquad
\sigma(e^+e^-\to\nu_e\bar N)=
\frac{1-P_{e^-}}{2}\sigma_0 .
\]
Neglecting electron-mass effects and possible CP-violating contributions,
the two charge-conjugate processes have identical fully polarized rates.
Consequently, their inclusive contribution satisfies
\[
\sigma(e^+e^-\to\nu_e\bar N+\bar\nu_eN)=\sigma_0,
\]
which explains the polarization independence of the inclusive neutral-VLL
production rate observed in the numerical analysis.

In contrast, both electron helicity components contribute to the charged
channel,
\[
\sigma_E(P_{e^-})=
\frac{1-P_{e^-}}{2}\sigma_E^L+
\frac{1+P_{e^-}}{2}\sigma_E^R .
\]
For the doublet-VLL scenario considered here, the right-handed contribution
is larger than the left-handed one,
$\sigma_E^R>\sigma_E^L$.
Therefore, a positive electron-beam polarization enhances the
charged-VLL production rate.
\subsection{Signal signatures and simulation}
\label{sec:signal-signatures}

We focus on the hadronic decay modes of the EW bosons,
\begin{equation}
W^\pm\to q\bar{q}', \qquad
h\to b\bar{b}, \qquad
Z\to q\bar{q},
\end{equation}
which provide sizable branching ratios and allow efficient
reconstruction of the heavy VLL decay products. At the detector level,
the hadronic $W$ and $Z$ decays are reconstructed as two light-jet systems,
which we denote by $2j$ throughout the following analysis.
Both charge-conjugate processes are included in the signal samples. For the charged-VLL channels, the
beam-polarization configuration
\[
(P_{e^-},P_{e^+})=(0.8,0)
\]
is applied to both signal and background samples.

The three signal topologies considered in this analysis are as follows.

\begin{itemize}

\item \textbf{Neutral-VLL signal:}
\begin{equation}
e^+e^- \to \nu_e\bar{N}+\bar{\nu}_e N
\to e^\pm+2j+\slashed{E}_T,
\end{equation}
where the neutral VLL decays through
$N\to W^-e^+$ and $\bar{N}\to W^+e^-$,
followed by the hadronic decay
$W^\pm\to q\bar{q}'$, reconstructed as a two light-jet system.

\item \textbf{Charged-VLL Higgs mode:}
\begin{equation}
e^+e^- \to e^\pm E^\mp
\to e^+e^-+b\bar{b},
\end{equation}
with $E^\pm\to e^\pm h$ and $h\to b\bar{b}$.
At the detector level, the two $b$ quarks are reconstructed as two
$b$-tagged jets, and the corresponding signature is denoted by
$e^+e^-+2b$.

\item \textbf{Charged-VLL $Z$ mode:}
\begin{equation}
e^+e^- \to e^\pm E^\mp
\to e^+e^-+q\bar{q},
\end{equation}
with $E^\pm\to e^\pm Z$ and $Z\to q\bar{q}$.
The hadronic $Z$ decay is reconstructed as a two light-jet system, resulting
in the detector-level signature
$e^+e^-+2j$.

\end{itemize}

The VLL model is implemented in
\texttt{FeynRules}~\cite{Alloul:2013bka} and exported in the Universal
FeynRules Output (UFO) format~\cite{Degrande:2011ua}, based on the
implementation of Ref.~\cite{Bhattiprolu:VLL-UFO}.
Signal and background events are generated with
\texttt{MadGraph5\_aMC@NLO}~\cite{Alwall:2014hca}. Parton showering and
hadronization are performed with \texttt{PYTHIA8}~\cite{Sjostrand:2014zea},
and detector effects are simulated using
\texttt{DELPHES}~\cite{deFavereau:2013fsa} with the CLIC detector card
\cite{Leogrande:2019qbe}. Initial-state radiation and beamstrahlung effects
are included using the 3-TeV CLIC beam spectrum.
Jets are reconstructed using the Valencia Linear Collider (VLC) algorithm
\cite{Boronat:2014hva,Boronat:2016tgd} in exclusive mode with two jets
($N=2$) and a radius parameter of $R=0.5$, optimized for the resolved
reconstruction of the hadronic boson decays.\footnote{A complementary fat-jet analysis is presented in
Appendix~\ref{app:fatjet} to investigate the impact of boosted-boson
reconstruction in the high-mass VLL regime.}
The $b$-tagging and mistagging efficiencies are modeled according to the
default CLIC parametrization implemented in \texttt{DELPHES}, including
their dependence on jet kinematics. Event reconstruction and the cut-based
analysis are performed with \texttt{MadAnalysis5}~\cite{Conte:2012fm}.

The baseline object-selection requirements are:
\begin{equation}
p_T^\ell>10~\mathrm{GeV},\qquad
|\eta_\ell|<2.5,\qquad
p_T^j>20~\mathrm{GeV},\qquad
|\eta_j|<5,
\end{equation}
where $\ell$ and $j$ denote reconstructed charged leptons and jets,
respectively.

\subsubsection{Signal 1 (neutral $N$)}
\label{sec:signal1}

\begin{figure*}[ht]
\begin{center}
\centerline{\hspace{3.0cm}\epsfxsize=10cm\epsffile{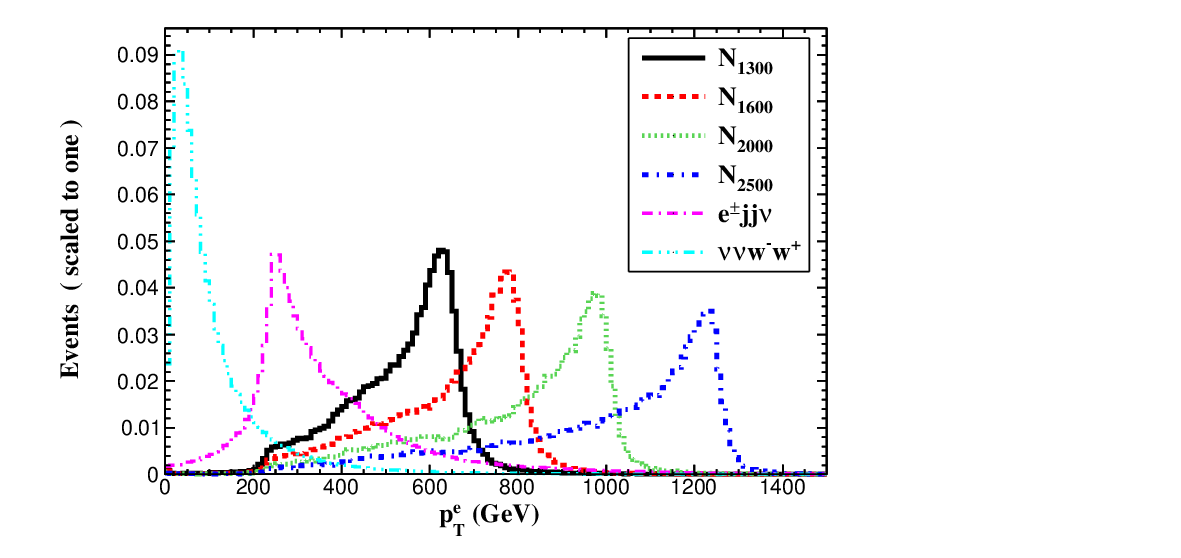}
\hspace{-3.0cm}\epsfxsize=10cm\epsffile{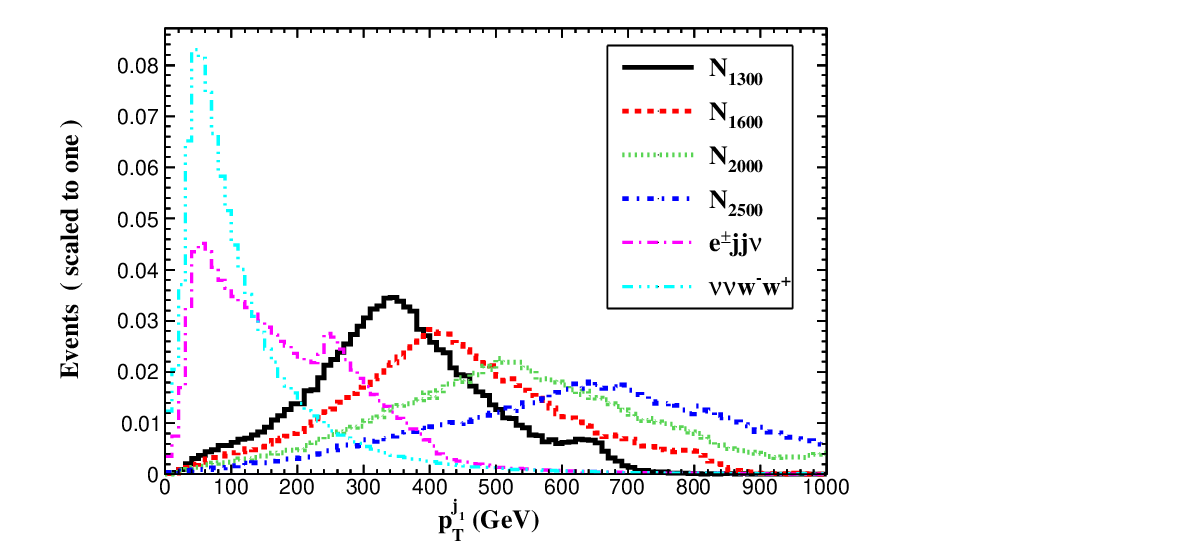}}
\centerline{\hspace{3.0cm}\epsfxsize=10cm\epsffile{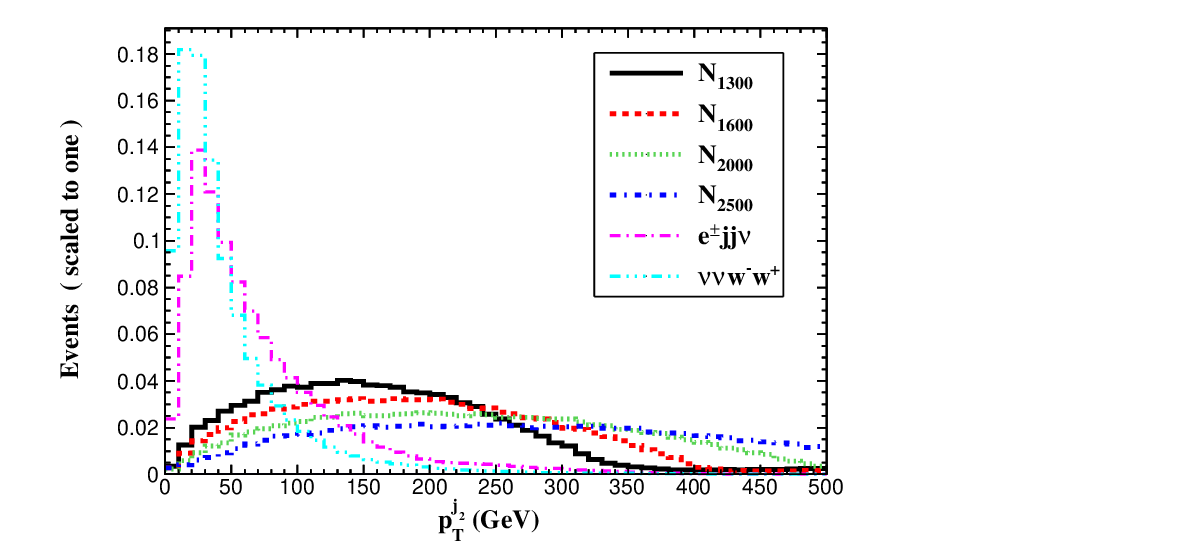}
\hspace{-3.0cm}\epsfxsize=10cm\epsffile{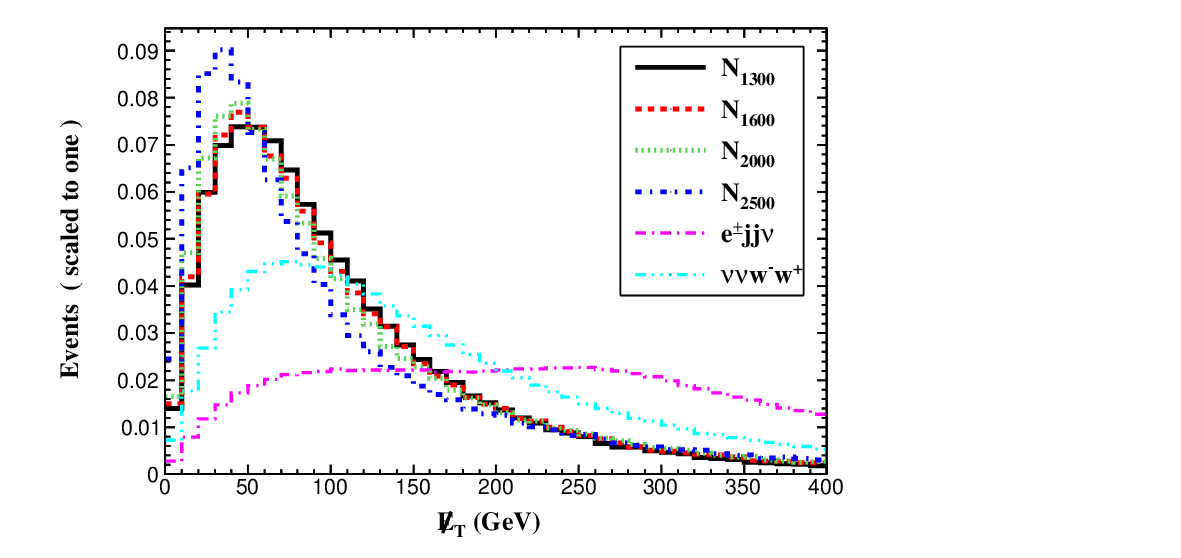}}
\caption{Normalized distributions of (a) $p_T^e$, (b) $p_T^{j_1}$,
(c) $p_T^{j_2}$, and (d) $\slashed{E}_T$ for Signal~1 and the dominant SM
backgrounds at $\sqrt{s}=3$~TeV.}
\label{fig:distribution1}
\end{center}
\end{figure*}

For the $e^\pm+2j+\slashed{E}_T$ final state, the dominant SM backgrounds
are
$e^+e^-\to e^\pm jj\nu$
and
$e^+e^-\to\nu\bar{\nu}W^+W^-$,
where the latter contributes when one $W$ boson decays leptonically and the
other hadronically. The signal is characterized by a high-energy isolated
electron from the heavy-VLL decay, a hadronically reconstructed $W$ boson,
and missing transverse momentum carried by the associated neutrino in the
production process.

Figure~\ref{fig:distribution1} shows the normalized distributions of
$p_T^e$, $p_T^{j_1}$, $p_T^{j_2}$, and $\slashed{E}_T$ for representative
signal benchmarks and the dominant backgrounds. The signal exhibits harder
electron and jet transverse-momentum spectra, while the missing transverse
momentum distribution peaks at lower values. These features motivate the
following selection criteria.

The cuts are applied sequentially.
\begin{itemize}
\item[\textbf{Cut 1:}]
Exactly one isolated electron is required. We define two predefined
signal regions: SR-I with $p_T^e>500$~GeV and SR-II with
$p_T^e>700$~GeV. SR-I is used for the mass hypotheses
$M_N\leq1600$~GeV, while SR-II is employed for $M_N>1600$~GeV.
The tighter requirement in SR-II exploits the harder electron spectrum
expected from the decay of a heavier neutral VLL.
\item \textbf{Cut 2:}
The two reconstructed jets are required to satisfy
$p_T^{j_1}>\SI{300}{GeV}$ and
$p_T^{j_2}>\SI{100}{GeV}$.
In addition, the dijet invariant mass must satisfy
\[
|M_{jj}-M_W|<\SI{20}{GeV},
\]
selecting events compatible with a hadronically decaying $W$ boson.
\item \textbf{Cut 3:}
The missing transverse momentum is required to satisfy
\[
\slashed{E}_T<\SI{100}{GeV}.
\]
This upper requirement exploits the relatively soft missing-momentum spectrum
of the signal and suppresses backgrounds containing larger sources of missing
momentum.
\end{itemize}

Before presenting the individual selection requirements, we emphasize
that the use of mass-dependent signal regions does not imply prior
knowledge of the physical VLL mass. Rather, the analysis should be
interpreted as a scan over different VLL-mass hypotheses. For each
tested mass hypothesis, a predefined signal region is employed and the
corresponding expected sensitivity is evaluated. This is analogous to
the standard procedure adopted in searches for heavy new particles,
where different signal regions may be optimized for different assumed
signal masses. No event-by-event knowledge of the VLL mass is therefore
assumed in defining the selections.

\begin{table}[ht]
\centering
\footnotesize
\setlength{\tabcolsep}{3pt}
\caption{
Cutflow for Signal~1. The entries denote cross sections in fb for the
$e^\pm+2j+\slashed{E}_T$ final state with $\epsilon=0.1$.
SR-I and SR-II correspond to the predefined requirements
$p_T^e>500$~GeV and $p_T^e>700$~GeV, respectively.
SR-I is used for the mass hypotheses $M_N\leq1600$~GeV, while SR-II is
used for $M_N>1600$~GeV.
}
\label{cutflow1}
\begin{tabular}{@{}l cccc cccc@{}}
\toprule[1.5pt]
\multirow{2}{*}{Cuts} & \multicolumn{4}{c}{Signal ($M_N$ in GeV)} &
\multicolumn{4}{c}{Background} \\
\cmidrule(lr){2-5} \cmidrule(lr){6-9}
& 1300 & 1600 & 2000 & 2500
& $ejj\nu$ (I) & $ejj\nu$ (II) & $\nu\bar{\nu}W^+W^-$ (I) & $\nu\bar{\nu}W^+W^-$ (II) \\
\midrule[0.8pt]
Basic & 8.79 & 4.91 & 2.29 & 0.69 & 479 & 479 & 17 & 17 \\
Cut 1 & 4.66 & 3.32 & 1.43 & 0.50 & 62.2 & 23.5 & 0.25 & 0.10 \\
Cut 2 & 2.95 & 2.45 & 1.18 & 0.42 & 1.01 & 0.37 & 0.06 & 0.03 \\
Cut 3 & 1.93 & 1.52 & 0.72 & 0.27 & 0.207 & 0.048 & 0.008 & 0.004 \\
\bottomrule[1.5pt]
\end{tabular}
\end{table}

The corresponding cutflow is summarized in Table~\ref{cutflow1}. After the
full selection, the total background cross sections are reduced to
approximately $0.215$~fb and $0.052$~fb in regions I and II, respectively.
Thus, the combination of a hard-electron requirement, hadronic $W$
reconstruction, and an upper cut on the missing transverse momentum provides
efficient background rejection while retaining a sizable fraction of the
signal.

\subsubsection{Signal 2 (charged $E$, Higgs mode)}

\begin{figure*}[ht]
\begin{center}
\centerline{\hspace{3.0cm}\epsfxsize=10cm\epsffile{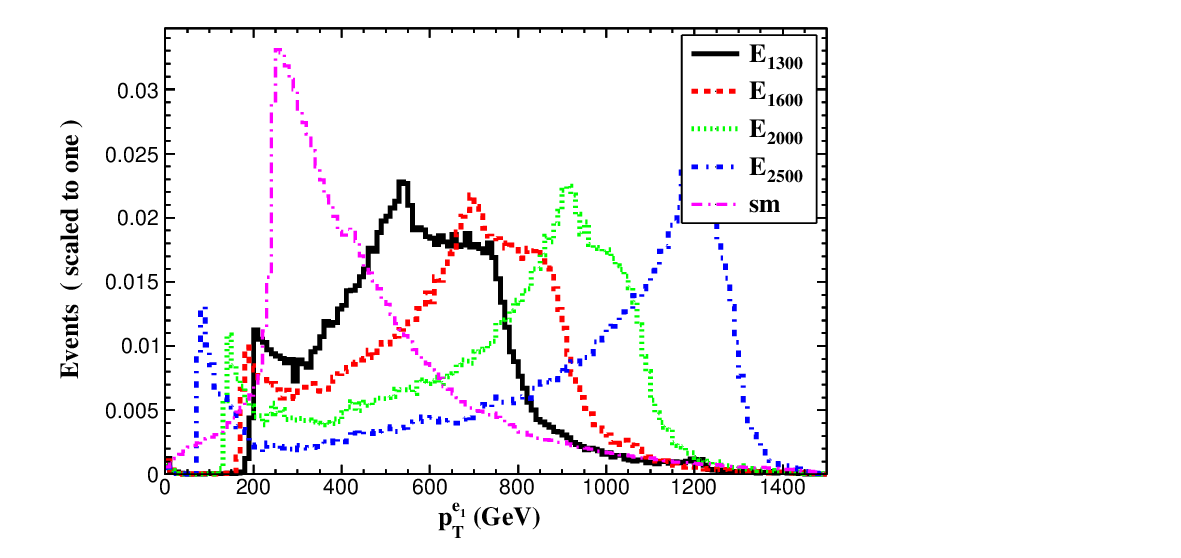}
\hspace{-3.0cm}\epsfxsize=10cm\epsffile{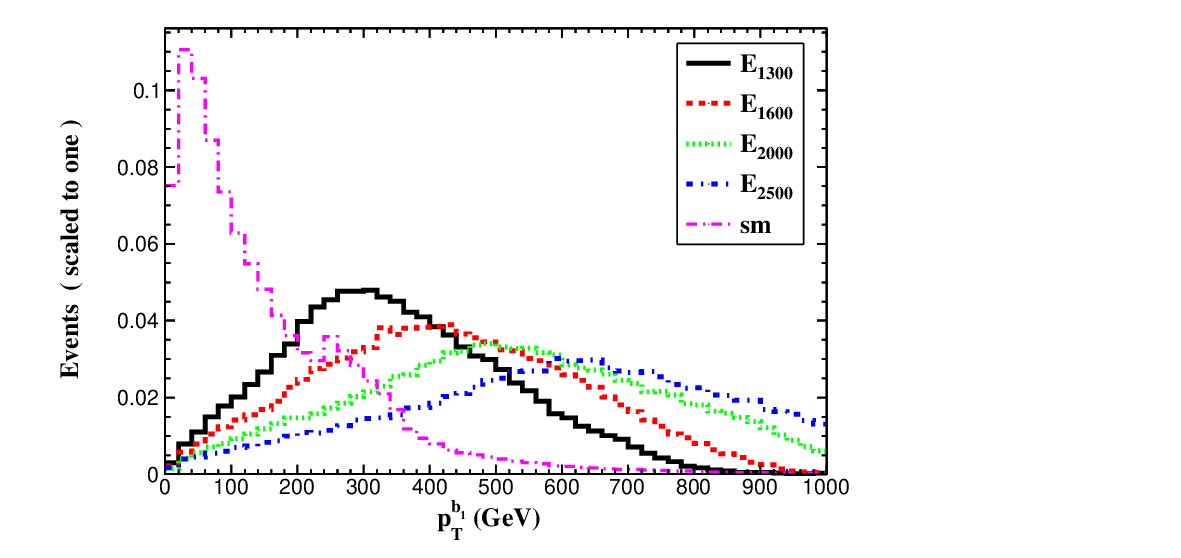}}
\centerline{\hspace{3.0cm}\epsfxsize=10cm\epsffile{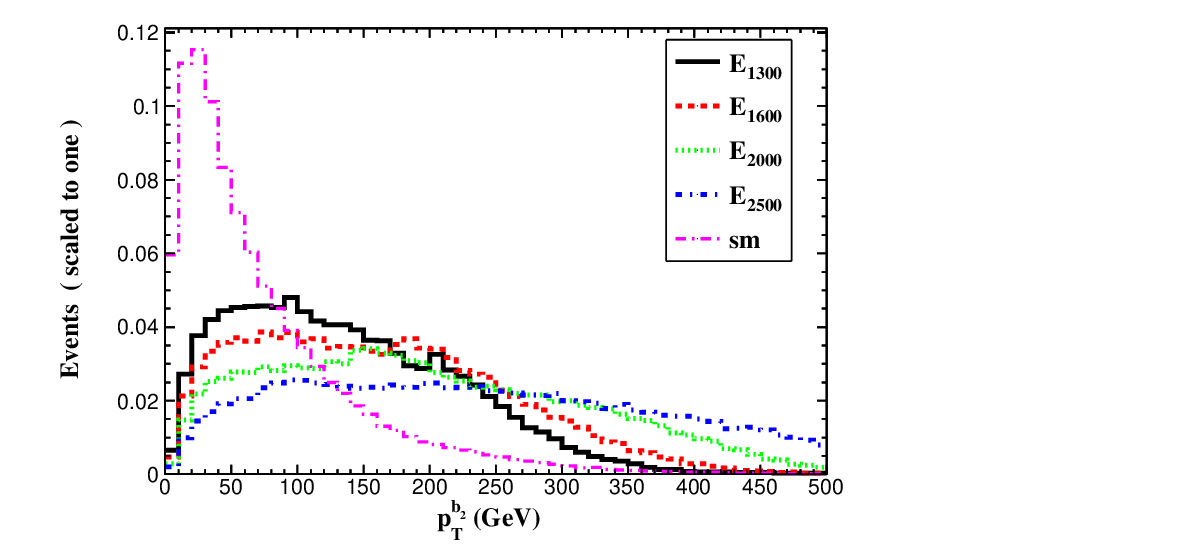}
\hspace{-3.0cm}\epsfxsize=10cm\epsffile{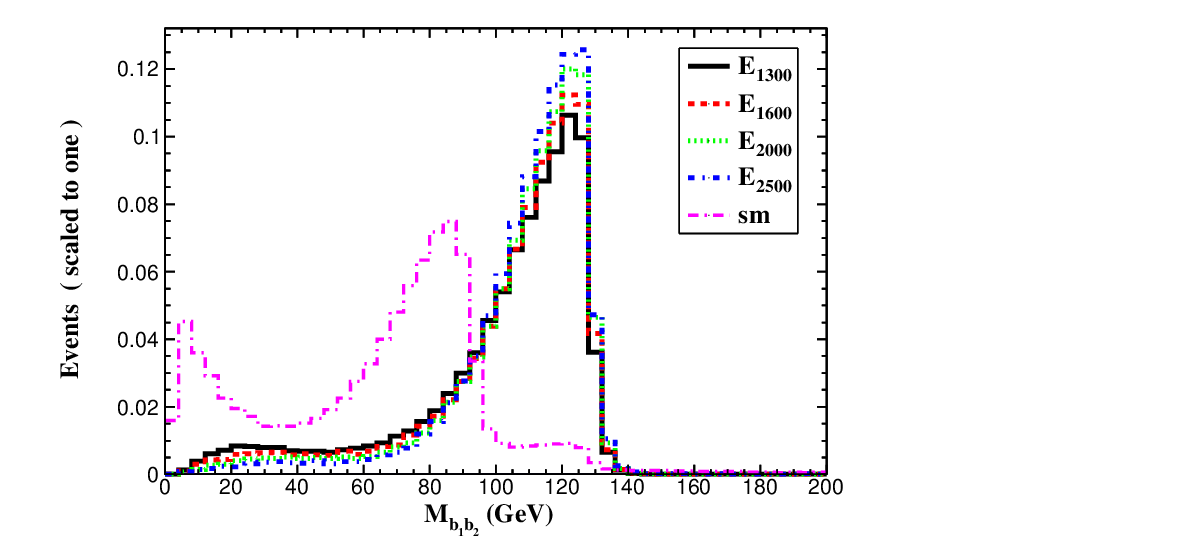}}
\centerline{\hspace{3.0cm}\epsfxsize=10cm\epsffile{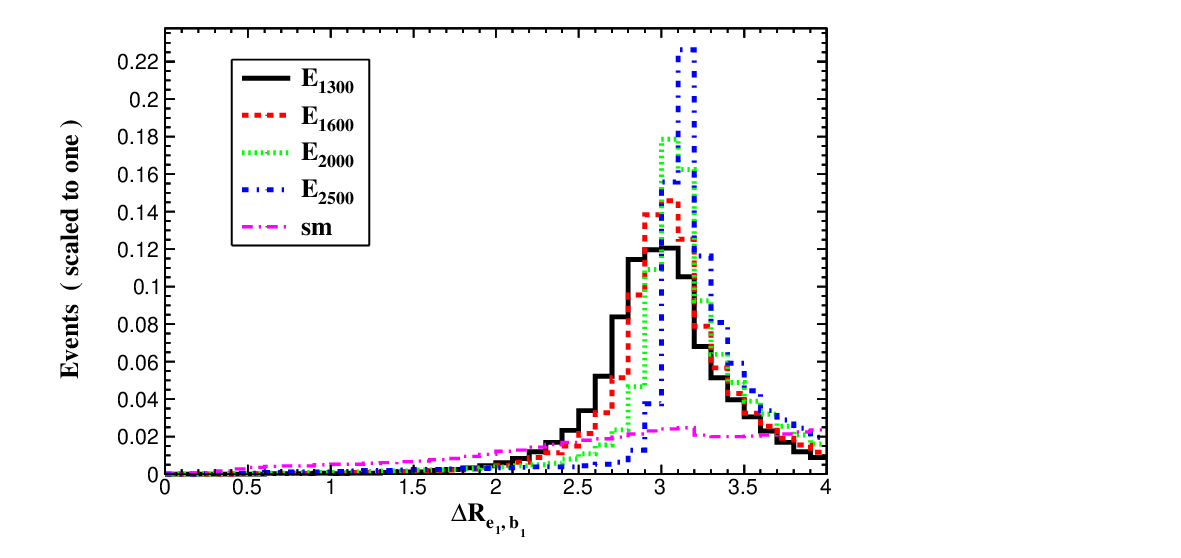}
\hspace{-3.0cm}\epsfxsize=10cm\epsffile{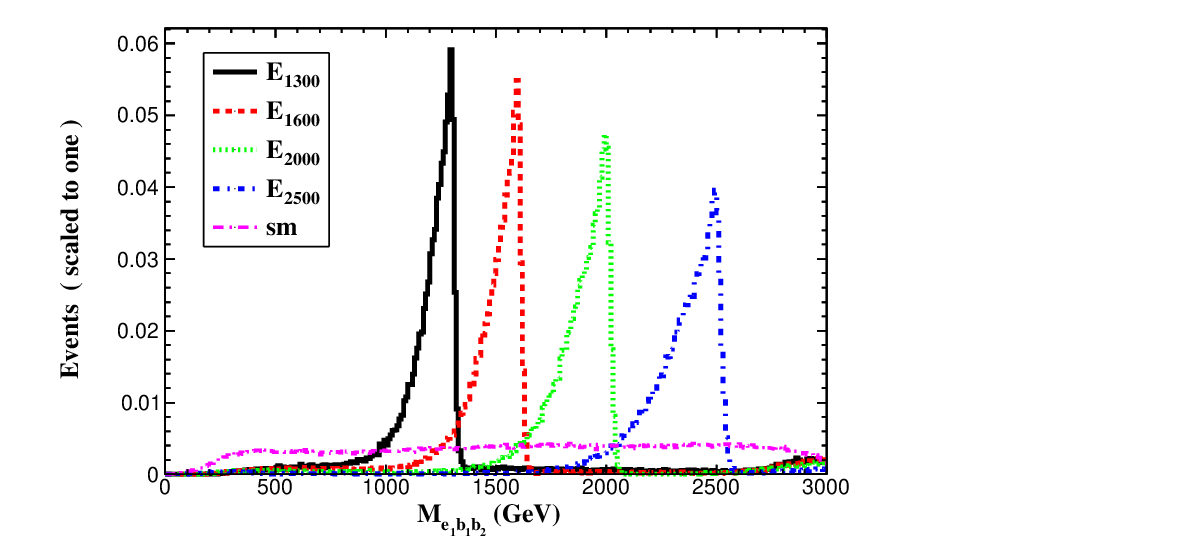}}
\caption{Normalized distributions of $p_T^{e_1}$, $p_T^{b_1}$,
$p_T^{b_2}$, $M_{b_1b_2}$, $\Delta R_{e_1b_1}$, and $M_{e_1b_1b_2}$
for Signal~2 and the dominant SM background at $\sqrt{s}=3$~TeV.}
\label{fig:distribution2}
\end{center}
\end{figure*}

For the Higgs mode,
$e^+e^-\to e^\pm E^\mp\to e^+e^-b\bar b$,
the dominant background is the inclusive
$e^+e^-\to e^+e^-jj$ process. Heavy-flavor production provides an
irreducible contribution, while light-flavor jets may enter through
mistagging.

Figure~\ref{fig:distribution2} presents the normalized distributions of
$p_T^{e_1}$, $p_T^{b_1}$, $p_T^{b_2}$, $M_{b_1b_2}$,
$\Delta R_{e_1b_1}$, and $M_{e_1b_1b_2}$.
The signal exhibits harder transverse-momentum spectra due to the decay of
a TeV-scale VLL. The $M_{b_1b_2}$ distribution is concentrated around the
Higgs-boson mass, while $M_{e_1b_1b_2}$ reconstructs the heavy-VLL mass
scale, providing effective discrimination against the SM background.

The cuts are again applied sequentially.
\begin{itemize}
\item \textbf{Cut 1:}
Exactly two isolated, oppositely charged electrons are required, with
\[
p_T^{e_1}>\SI{500}{GeV},
\qquad
p_T^{e_2}>\SI{10}{GeV}.
\]
The hard leading-electron requirement targets the electron from the heavy-VLL
decay, whereas the softer electron is predominantly associated with the
production process.
\item \textbf{Cut 2:}
Two $b$-tagged jets are required to satisfy
\[
p_T^{b_1}>\SI{200}{GeV},
\qquad
p_T^{b_2}>\SI{100}{GeV}.
\]
Their invariant mass must further satisfy
\[
|M_{b_1b_2}-M_h|<\SI{20}{GeV},
\]
selecting candidates compatible with $h\to b\bar b$.
\item \textbf{Cut 3:}
The reconstructed VLL mass is required to satisfy
\[
M_{e_1b_1b_2}>\SI{1200}{GeV},
\]
together with
\[
2.5<\Delta R_{e_1b_1}<3.5.
\]
The invariant-mass requirement suppresses the lower-scale SM background,
while the angular requirement exploits the nearly back-to-back topology
inherited from the two-body decay $E^\pm\to e^\pm h$.
\end{itemize}

\begin{table}[htb]
\centering
\small
\caption{Cutflow for Signal~2,
$e^+e^-\to e^\pm E^\mp\to e^+e^-b\bar b$, assuming
$\varepsilon=0.1$. The entries denote cross sections in fb.}
\label{cutflow2}
\vspace{0.1cm}
\setlength{\tabcolsep}{5pt}
\begin{tabular}{@{}l cccc c@{}}
\toprule[1.5pt]
\multirow{2}{*}{Cuts} &
\multicolumn{4}{c}{Signal ($M_E$ in GeV)} & Background \\
\cmidrule(lr){2-5}
& 1300 & 1600 & 2000 & 2500 & $e^+e^-jj$ \\
\midrule[0.8pt]
Basic & 0.084 & 0.041 & 0.023 & 0.011 & 14.8 \\
Cut 1 & 0.040 & 0.023 & 0.015 & 0.007 & 4.06 \\
Cut 2 & 0.013 & 0.009 & 0.0066 & 0.0034 & 0.023 \\
Cut 3 & 0.010 & 0.0086 & 0.0059 & 0.0028 & 0.0116 \\
\bottomrule[1.5pt]
\end{tabular}
\end{table}

The cutflow results are summarized in Table~\ref{cutflow2}. After the full
selection, the background cross section is reduced from $14.8$~fb to
approximately $0.012$~fb. The surviving signal cross sections decrease from
$0.010$~fb at $M_E=1300$~GeV to $0.0028$~fb at $M_E=2500$~GeV.
Relative to the basic selection, the corresponding signal efficiencies are
approximately $12\%$, $21\%$, $26\%$, and $25\%$ for
$M_E=1300$, $1600$, $2000$, and $2500$~GeV, respectively. The strongest
background suppression arises from the combined effects of $b$ tagging,
Higgs-mass reconstruction, and the heavy-VLL kinematic requirements.

\subsubsection{Signal 3 (charged $E$, $Z$ mode)}

\begin{figure*}[ht]
\begin{center}
\centerline{\hspace{3.0cm}\epsfxsize=10cm\epsffile{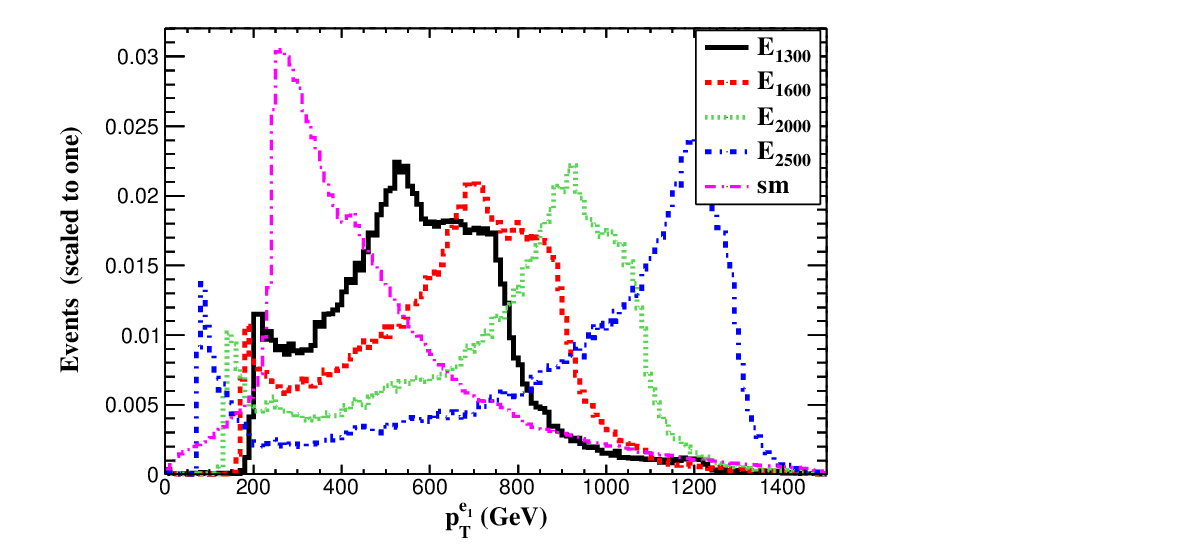}
\hspace{-3.0cm}\epsfxsize=10cm\epsffile{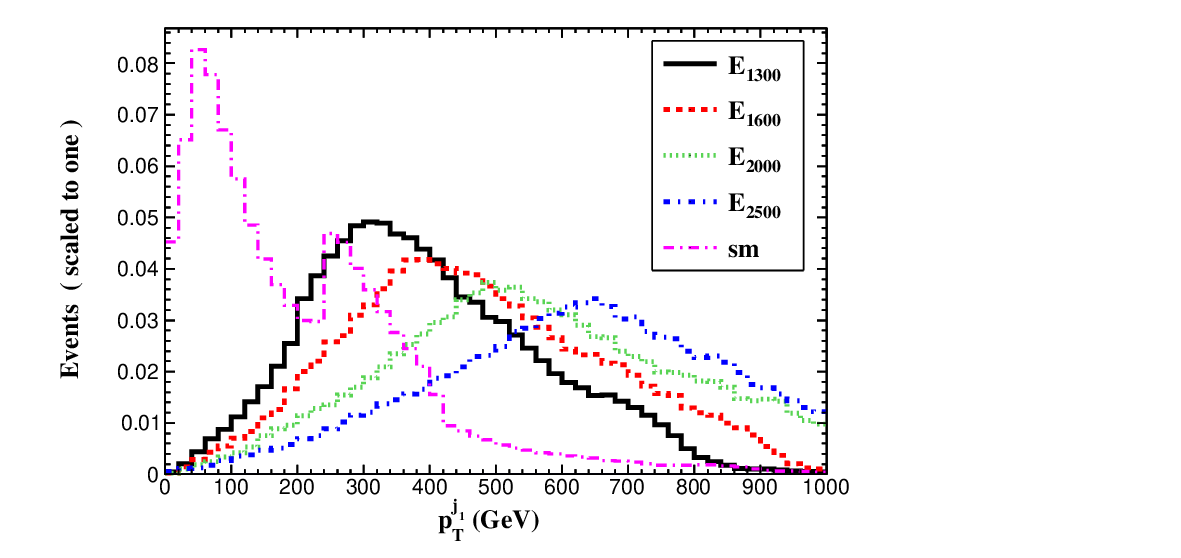}}
\centerline{\hspace{3.0cm}\epsfxsize=10cm\epsffile{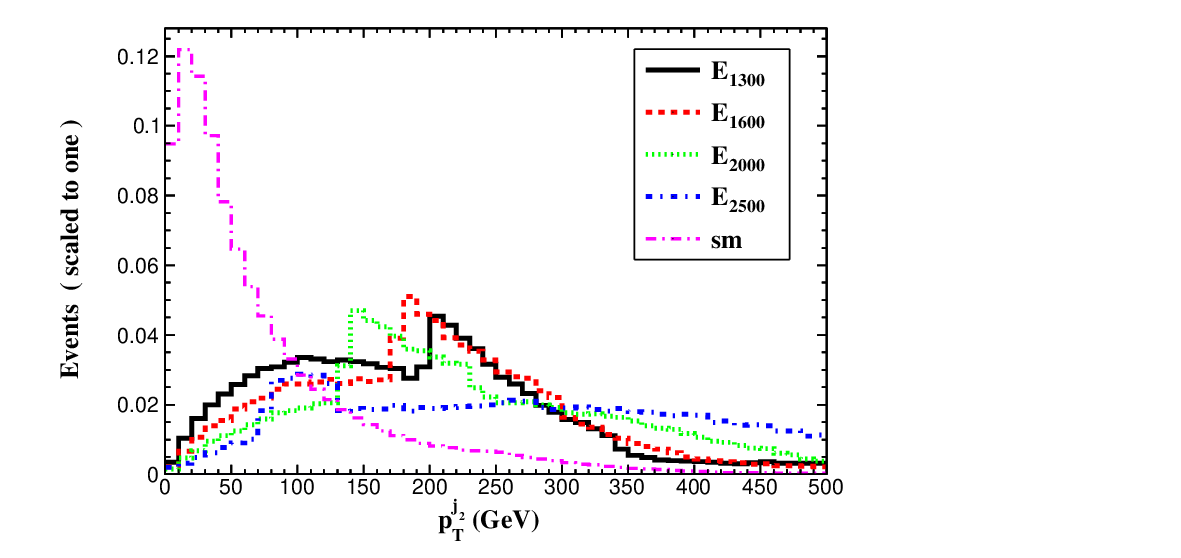}
\hspace{-3.0cm}\epsfxsize=10cm\epsffile{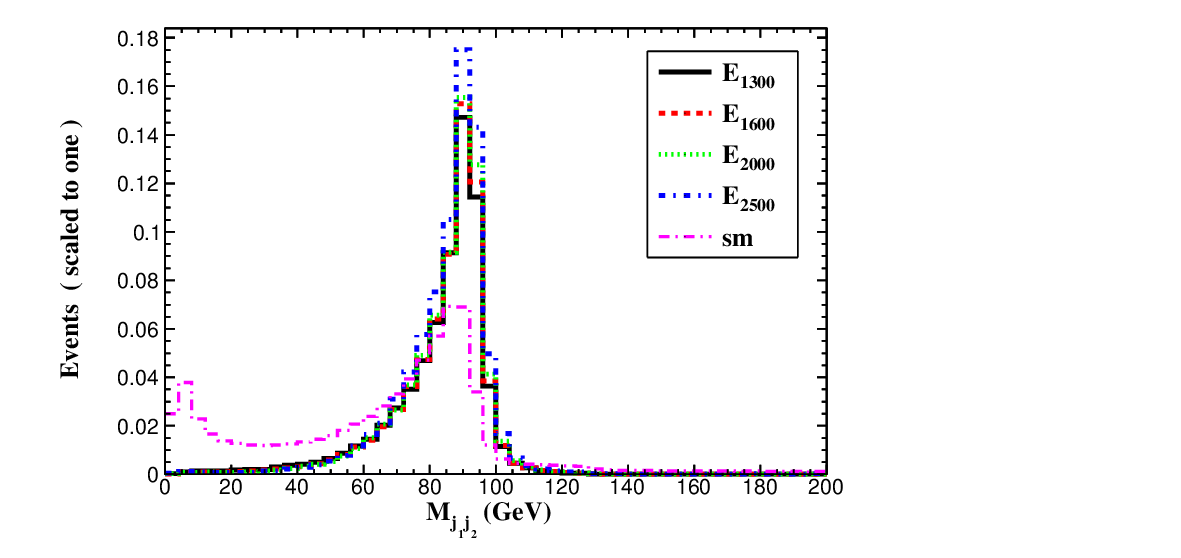}}
\centerline{\hspace{3.0cm}\epsfxsize=10cm\epsffile{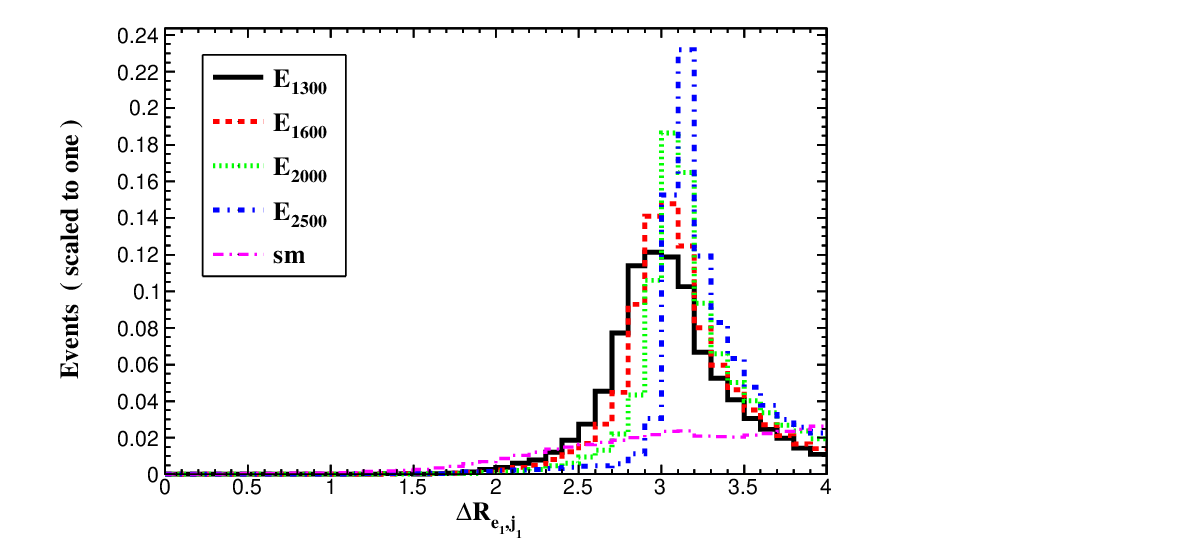}
\hspace{-3.0cm}\epsfxsize=10cm\epsffile{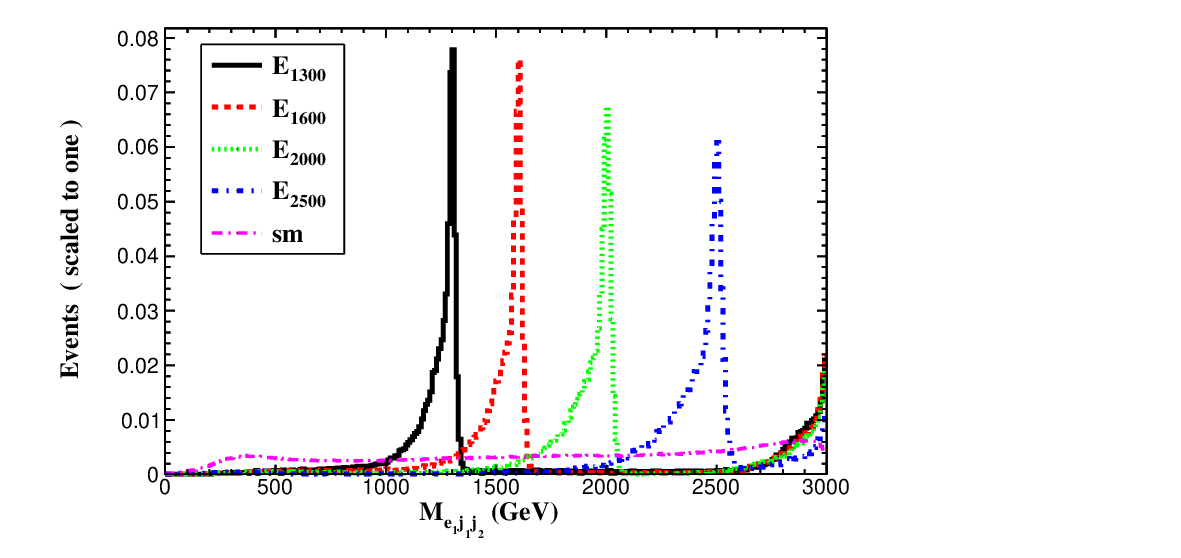}}
\caption{Normalized distributions of $p_T^{e_1}$, $p_T^{j_1}$,
$p_T^{j_2}$, $M_{j_1j_2}$, $\Delta R_{e_1j_1}$, and
$M_{e_1j_1j_2}$ for Signal~3 and the dominant SM background at
$\sqrt{s}=3$~TeV.}
\label{fig:distribution3}
\end{center}
\end{figure*}

For the hadronic $Z$ mode, $e^+e^-\to e^\pm E^\mp\to e^+e^-q\bar q$,
the dominant background is again the inclusive
$e^+e^-\to e^+e^-jj$ process. In contrast to Signal 2, no flavor-tagging requirement is imposed because the hadronic $Z$ decay is reconstructed inclusively.
Consequently, a larger continuum contribution remains after selection.

Figure~\ref{fig:distribution3} shows the normalized distributions of the
leading-electron transverse momentum $p_T^{e_1}$, the transverse momenta of
the leading and subleading jets $p_T^{j_1}$ and $p_T^{j_2}$, the dijet
invariant mass $M_{j_1j_2}$, the angular separation
$\Delta R_{e_1j_1}$, and the reconstructed VLL mass
$M_{e_1j_1j_2}$. The signal exhibits harder transverse-momentum spectra and
larger three-body invariant masses than the background. In addition,
$M_{j_1j_2}$ is concentrated around the $Z$-boson mass, providing a useful
handle against the continuum background.

The selection follows that of Signal~2, with the Higgs reconstruction
replaced by hadronic $Z$ reconstruction.
\begin{itemize}
\item \textbf{Cut 1:}
Exactly two isolated, oppositely charged electrons are required, with
\[
p_T^{e_1}>\SI{500}{GeV},
\qquad
p_T^{e_2}>\SI{10}{GeV}.
\]
\item \textbf{Cut 2:}
Two reconstructed jets are required to satisfy
\[
p_T^{j_1}>\SI{200}{GeV},
\qquad
p_T^{j_2}>\SI{100}{GeV}.
\]
Their invariant mass must satisfy
\[
|M_{j_1j_2}-M_Z|<\SI{15}{GeV},
\]
selecting candidates compatible with a hadronically decaying $Z$ boson.
\item \textbf{Cut 3:}
The reconstructed VLL mass is required to satisfy
\[
M_{e_1j_1j_2}>\SI{1200}{GeV},
\]
together with
\[
2.5<\Delta R_{e_1j_1}<3.5.
\]
The invariant-mass requirement suppresses the lower-scale SM background,
while the angular requirement exploits the nearly back-to-back topology
inherited from the two-body decay $E^\pm\to e^\pm Z$.

\end{itemize}

We note that, unlike the two signal regions introduced for Signal~1,
the selection requirements for Signals~2 and~3 are kept fixed over the
full VLL-mass range considered. In particular, the reconstructed
three-body invariant-mass requirement is imposed as a common lower
threshold rather than as a mass window centered on an assumed value of
$M_E$.

\begin{table}[htb]
\centering
\small
\caption{Cutflow for Signal~3,
$e^+e^-\to e^\pm E^\mp\to e^+e^-q\bar q$, assuming
$\varepsilon=0.1$. The entries denote cross sections in fb.}
\label{cutflow3}
\vspace{0.1cm}
\setlength{\tabcolsep}{5pt}
\begin{tabular}{@{}l cccc c@{}}
\toprule[1.5pt]
\multirow{2}{*}{Cuts} &
\multicolumn{4}{c}{Signal ($M_E$ in GeV)} & Background \\
\cmidrule(lr){2-5}
& 1300 & 1600 & 2000 & 2500 & \(e^+e^-jj\) \\
\midrule[0.8pt]
Basic & 0.16 & 0.096 & 0.054 & 0.025 & 14.8 \\
Cut 1 & 0.075 & 0.055 & 0.034 & 0.017 & 4.07 \\
Cut 2 & 0.050 & 0.039 & 0.024 & 0.012 & 0.78 \\
Cut 3 & 0.042 & 0.036 & 0.021 & 0.0094 & 0.255 \\
\bottomrule[1.5pt]
\end{tabular}
\end{table}

The cutflow results are summarized in Table~\ref{cutflow3}. After the full
selection, the background cross section is reduced from $14.8$~fb to
approximately $0.255$~fb. The surviving signal cross sections decrease from
$0.042$~fb at $M_E=1300$~GeV to $0.0094$~fb at $M_E=2500$~GeV.
Relative to the basic selection, the corresponding signal efficiencies are
approximately $26\%$, $38\%$, $39\%$, and $38\%$ for
$M_E=1300$, $1600$, $2000$, and $2500$~GeV, respectively. The higher efficiency of the $Z$ channel mainly results from the absence of
the $b$-tagging requirement, which reduces signal loss despite the larger
continuum background.

\subsection{Statistical analysis}

The sensitivity to the VLL signals is quantified through the expected median
discovery and exclusion significances using the asymptotic formulae of
Ref.~\cite{Cowan:2010js},
\begin{equation}
\begin{aligned}
\mathcal{Z}_{\rm disc}
&=
\sqrt{
2\left[
(s+b)\ln\left(1+\frac{s}{b}\right)-s
\right]},
\\
\mathcal{Z}_{\rm excl}
&=
\sqrt{
2\left[
s-b\ln\left(1+\frac{s}{b}\right)
\right]},
\end{aligned}
\end{equation}
where $s$ and $b$ denote the expected signal and background yields,
respectively.

Since the single-production cross sections scale approximately as
$\varepsilon^2$, the signal yields for arbitrary mixing values are obtained
from the benchmark samples generated with $\varepsilon_0=0.1$ according to
\begin{equation}
s(M_L,\varepsilon)
=
s(M_L,\varepsilon_0)
\left(\frac{\varepsilon}{\varepsilon_0}\right)^2,
\qquad
\varepsilon_0=0.1 .
\end{equation}
For Signal~1, the sensitivity at each tested value of $M_N$ is evaluated
using the corresponding predefined signal region introduced in
Sec.~\ref{sec:signal1}. This corresponds to a scan over different
signal-mass hypotheses and does not require prior knowledge of the VLL mass.

The sensitivities are evaluated at $\sqrt{s}=3$~TeV for integrated
luminosities of $\mathcal{L}=1$, $3$, and $5~\mathrm{ab}^{-1}$.
The exclusion and discovery contours are defined by
$\mathcal{Z}_{\rm excl}=2$ and $\mathcal{Z}_{\rm disc}=5$, respectively.
The results correspond to statistical-only projections. Systematic
uncertainties associated with background modeling, detector effects,
luminosity determination, and beam polarization may modify the final
sensitivity and would require a dedicated experimental treatment.

\begin{figure*}[t!]
\begin{center}
\centerline{\hspace{1.0cm}\epsfxsize=8cm\epsffile{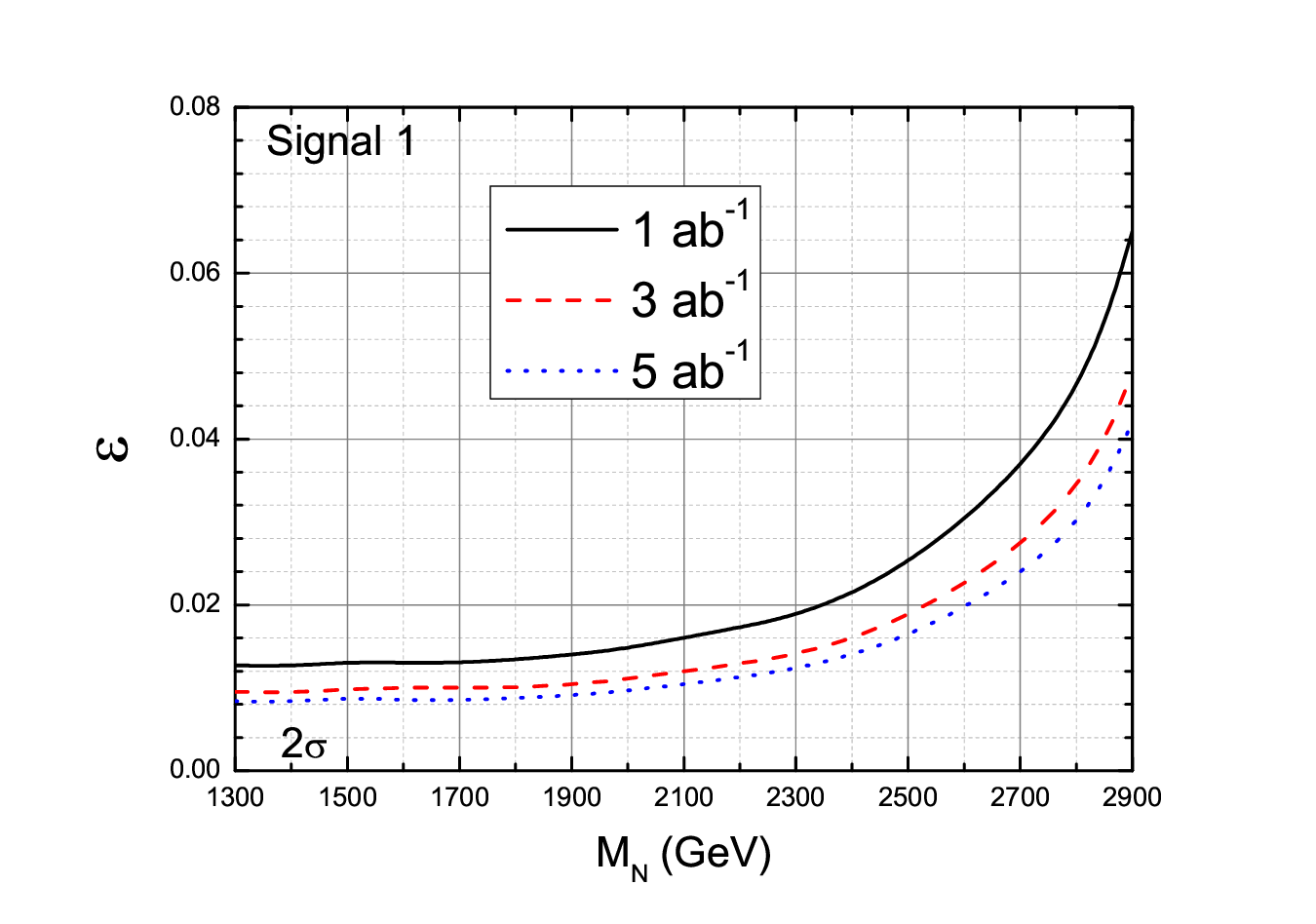}
\hspace{-1.0cm}\epsfxsize=8cm\epsffile{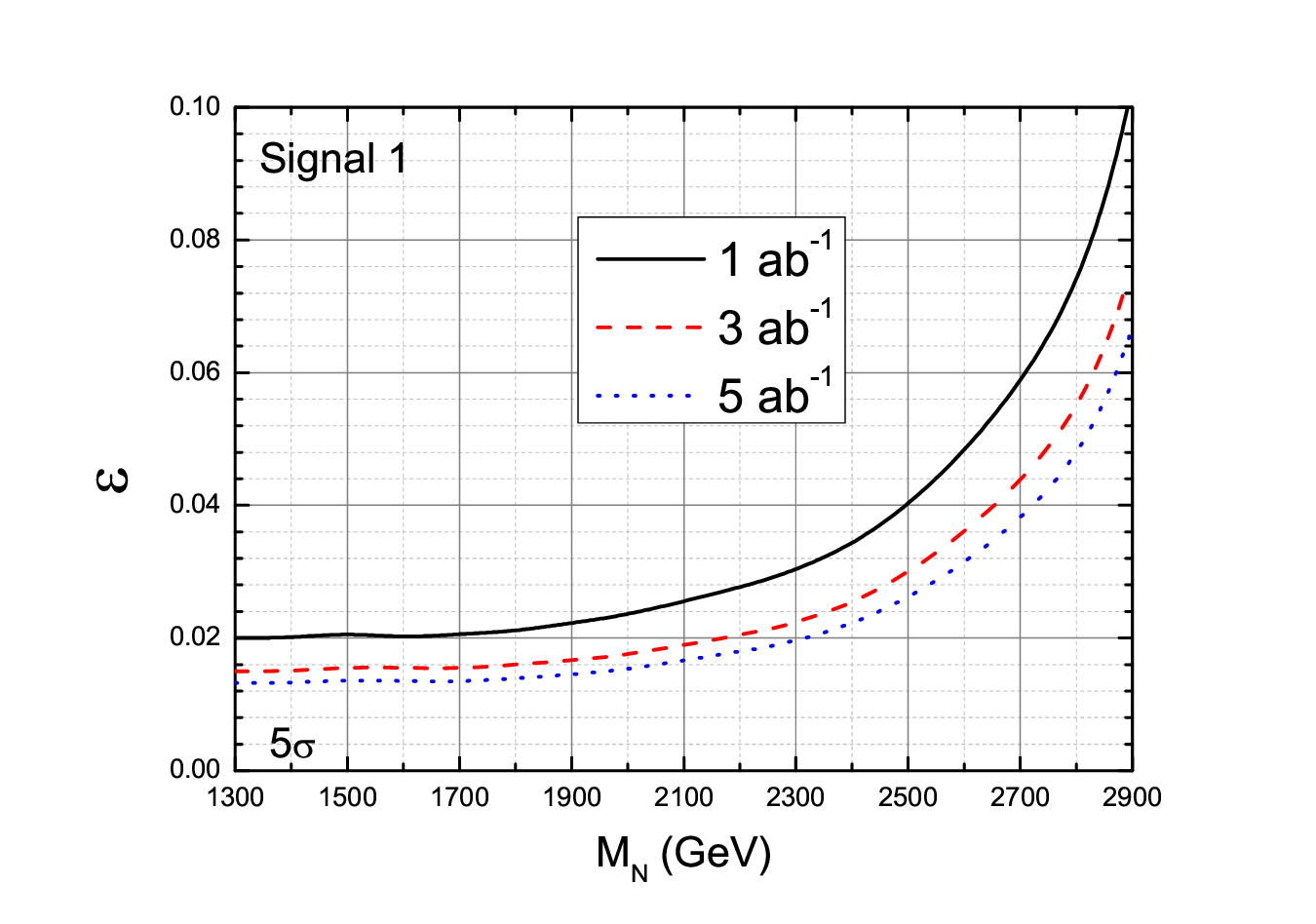}}
\centerline{\hspace{1.0cm}\epsfxsize=8cm\epsffile{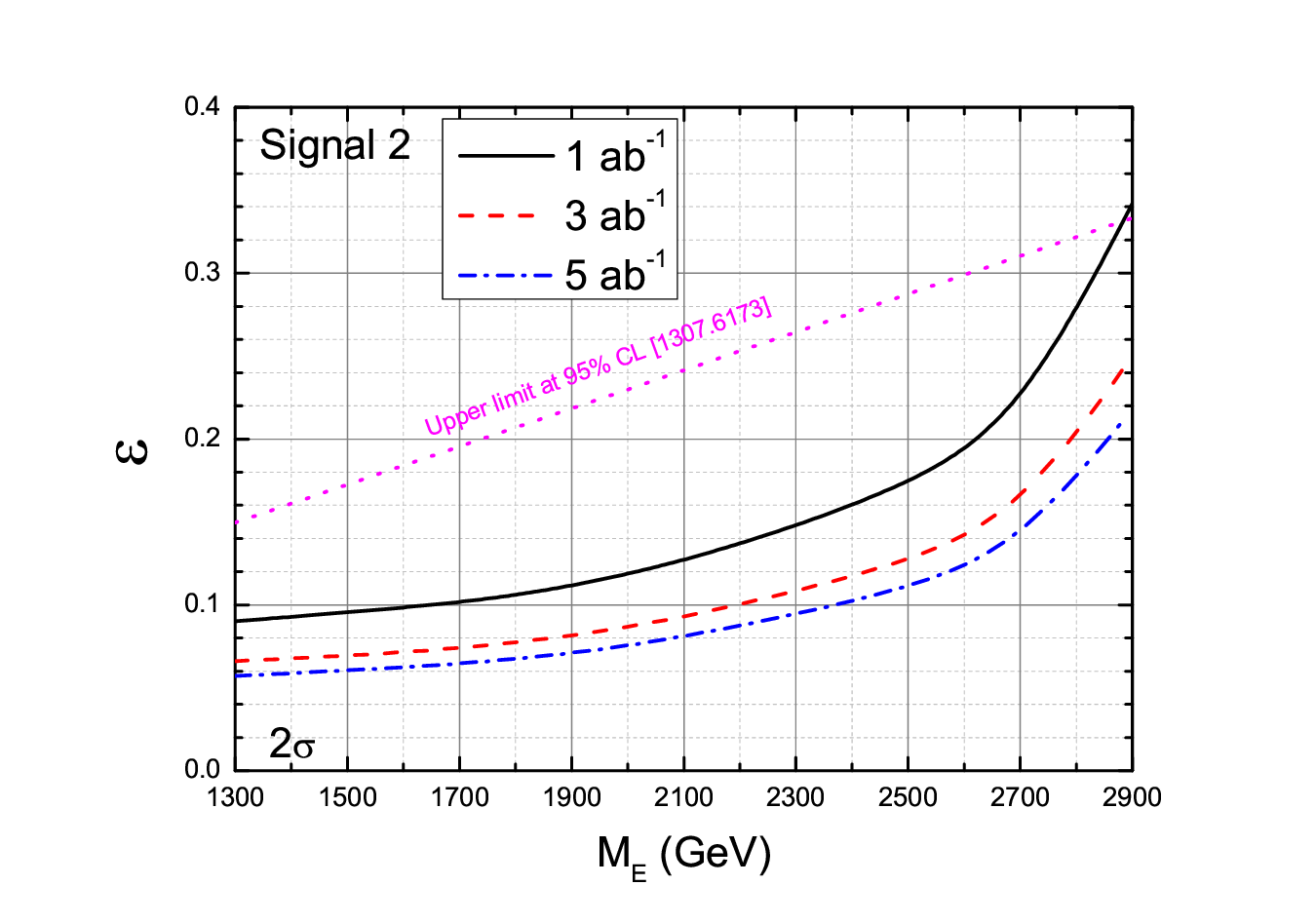}
\hspace{-1.0cm}\epsfxsize=8cm\epsffile{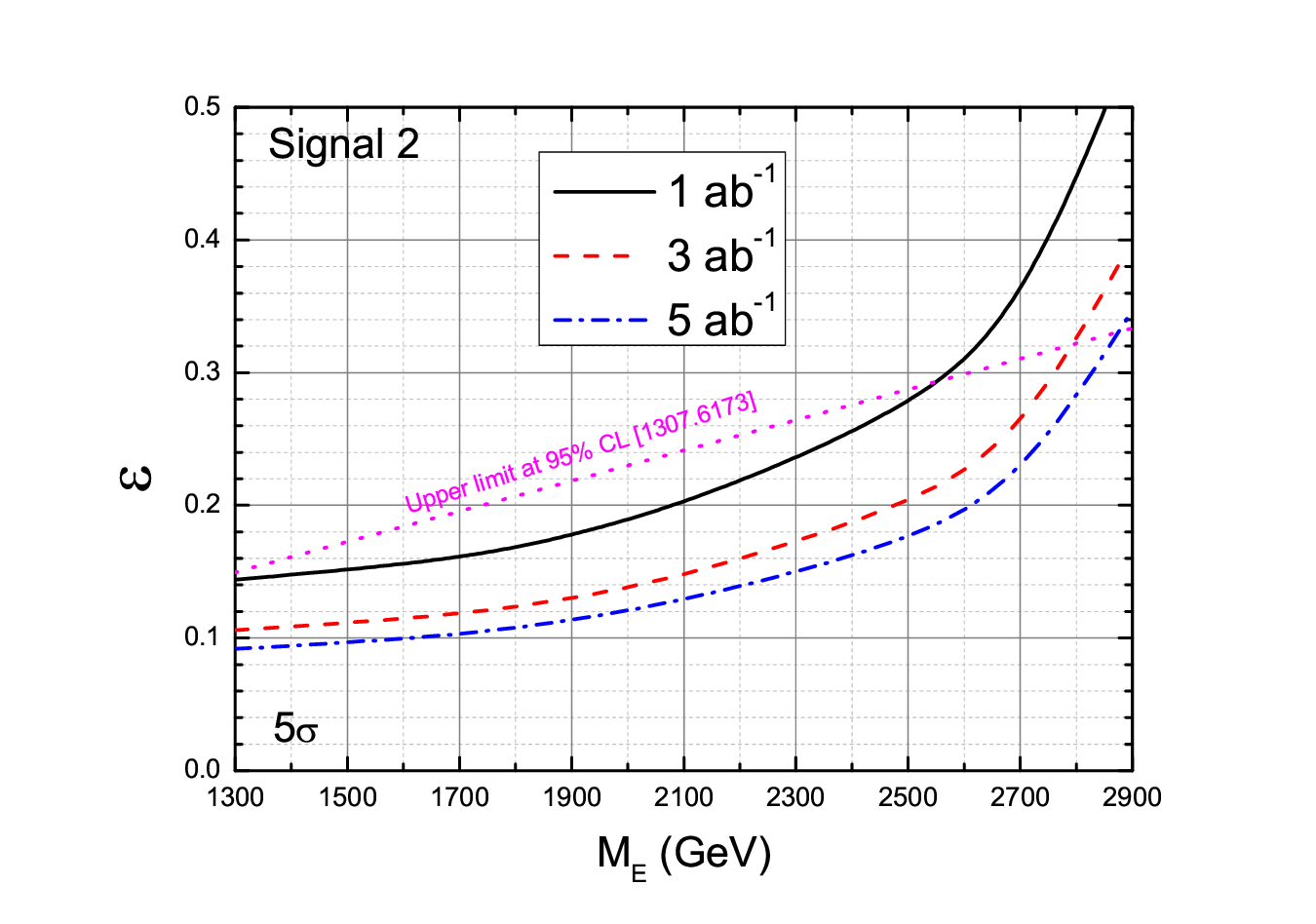}}
\centerline{\hspace{1.0cm}\epsfxsize=8cm\epsffile{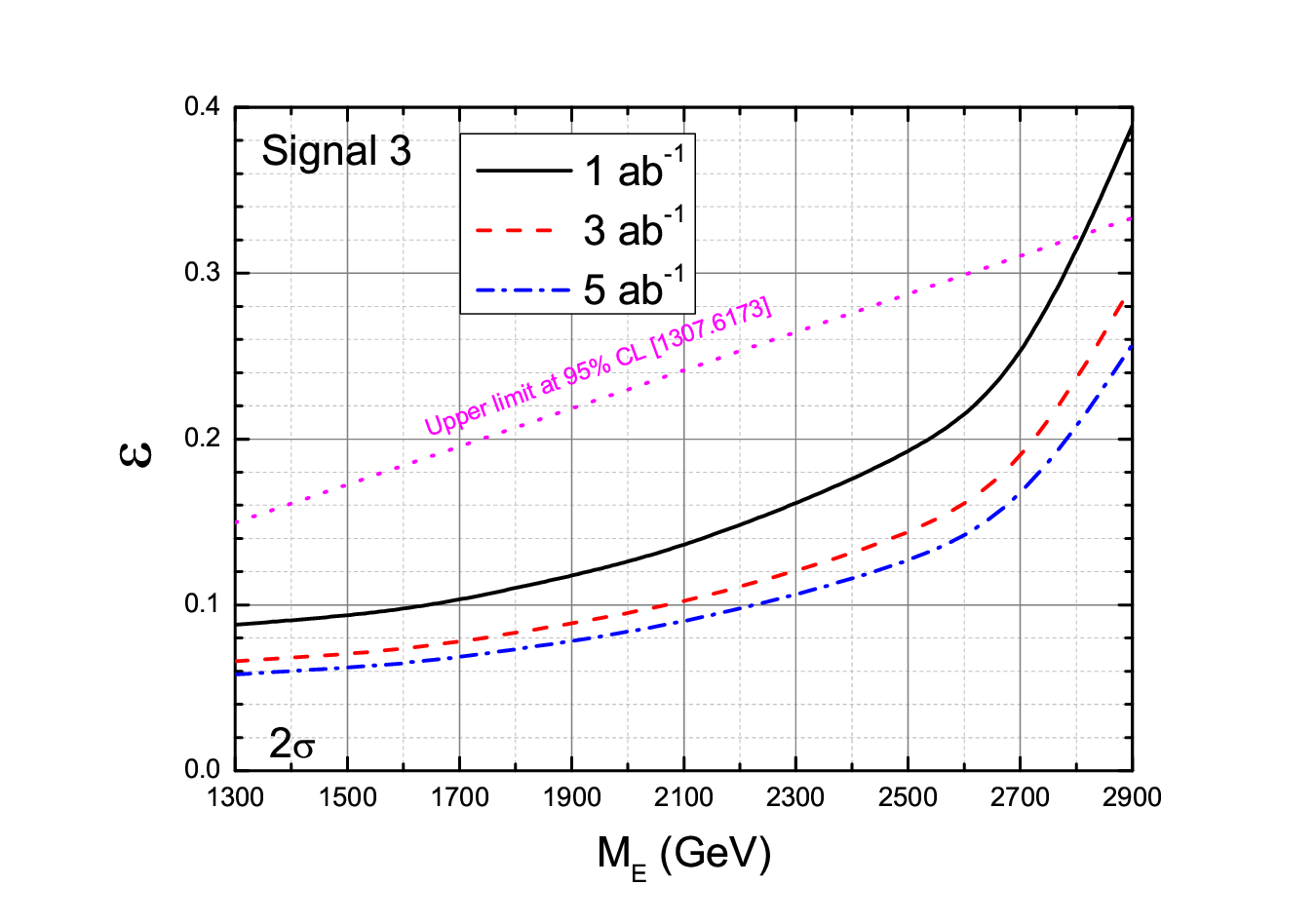}
\hspace{-1.0cm}\epsfxsize=8cm\epsffile{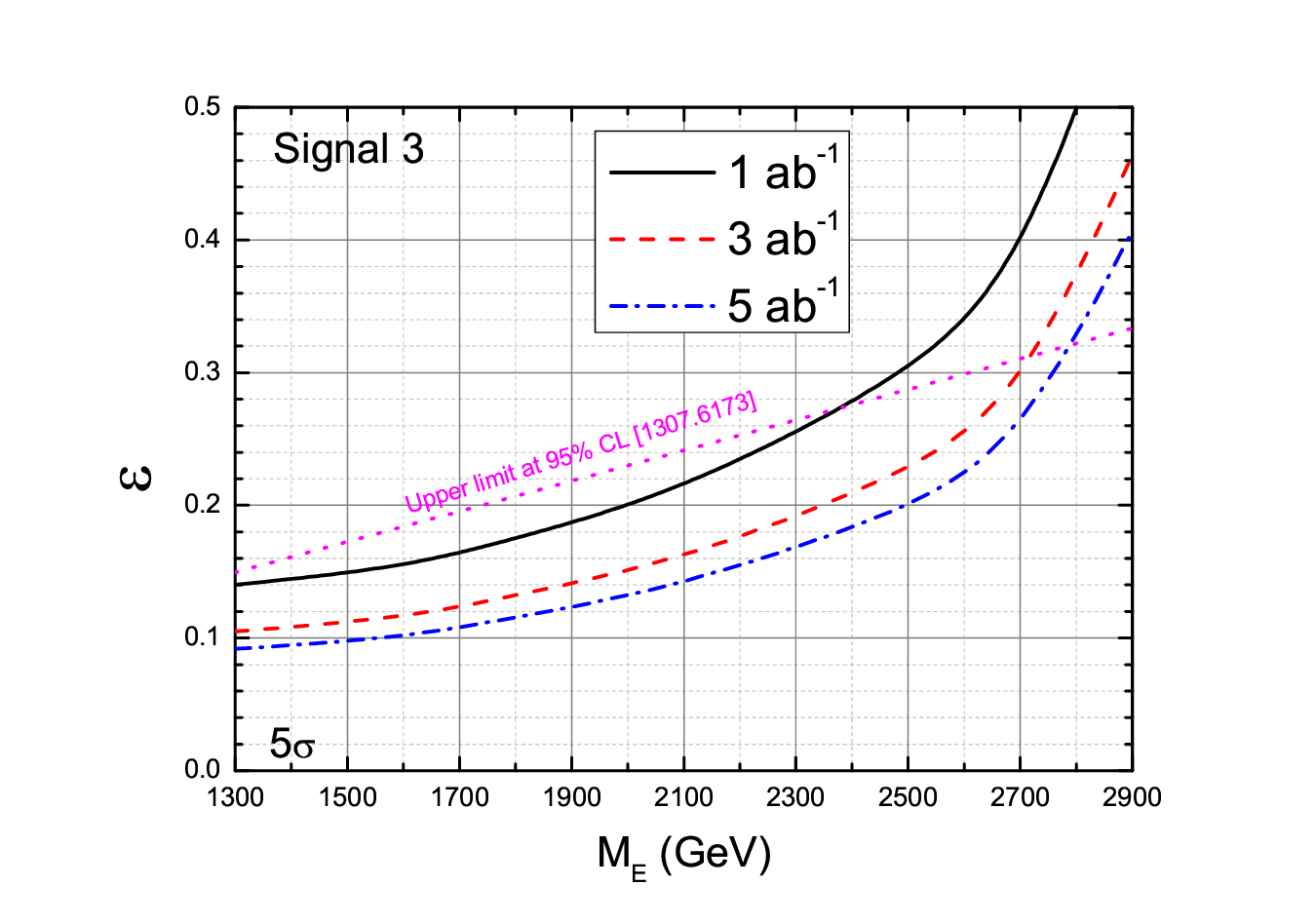}}
\caption{
Expected $2\sigma$ exclusion (left) and $5\sigma$ discovery (right) reaches for
Signal~1 (top), Signal~2 (middle), and Signal~3 (bottom) at $\sqrt{s}=3$~TeV
for $\mathcal{L}=1$, $3$, and $5~\mathrm{ab}^{-1}$. The dotted curves indicate
the EWPO upper bound for the first-generation doublet-VLL scenario,
$\sin\theta_\ell<0.02$~\cite{deBlas:2013gla}, and are shown for the charged
channels.
}
\label{fig-sensitivity}
\end{center}
\end{figure*}

Figure~\ref{fig-sensitivity} presents the expected $2\sigma$ exclusion and
$5\sigma$ discovery reaches in the $(M_L,\varepsilon)$ plane for the three
signal channels. For the charged-VLL channels, the dotted curves indicate
the indirect EWPO constraint,
$\sin\theta_\ell<0.02$, corresponding to
$\varepsilon<0.02M_L/v$.
For Signal~1, the EWPO boundary lies above the displayed parameter range
and is therefore not shown.

As expected, increasing the integrated luminosity improves the sensitivity
to smaller mixing parameters and extends the accessible VLL mass range.
Among the three channels, the neutral-VLL production channel provides the
highest sensitivity due to its larger production rate and cleaner final
state. For $\mathcal{L}=1~\mathrm{ab}^{-1}$, the expected $2\sigma$
exclusion sensitivity reaches
$\varepsilon\simeq0.013$ at $M_N=1300$~GeV and
$\varepsilon\simeq0.065$ at $M_N=2900$~GeV.
The corresponding $5\sigma$ discovery sensitivity ranges from approximately
$\varepsilon\simeq0.02$ to $0.10$ over the considered mass interval.

The charged-VLL channels exhibit weaker sensitivity, mainly because of their
smaller production cross sections. The $Z$-boson channel also retains a
larger residual background since no flavor-tagging requirement is imposed.
For $\mathcal{L}=1~\mathrm{ab}^{-1}$, the $2\sigma$ exclusion reaches are
approximately $\varepsilon\simeq0.09$--$0.34$ for Signal~2 and
$\varepsilon\simeq0.09$--$0.39$ for Signal~3, while the corresponding
$5\sigma$ discovery reaches are $\varepsilon\simeq0.14$--$0.55$ and
$\varepsilon\simeq0.14$--$0.62$, respectively.

At $\mathcal{L}=5~\mathrm{ab}^{-1}$, the $2\sigma$ exclusion reaches improve
to approximately $\varepsilon\simeq0.06$--$0.22$ and
$\varepsilon\simeq0.06$--$0.26$ for Signals~2 and~3, respectively,
while the corresponding $5\sigma$ discovery reaches become
$\varepsilon\simeq0.09$--$0.35$ and
$\varepsilon\simeq0.09$--$0.41$.

The comparison with the EWPO constraint demonstrates the complementarity
between direct collider searches and precision measurements. For Signal~1,
the CLIC sensitivity probes mixing parameters substantially below the
current indirect limit over the entire mass range considered. For the
charged channels, the direct-search sensitivity is comparable to or stronger
than the EWPO constraint at lower masses, while it deteriorates near the
kinematic limit due to the suppression of the single-production cross
section.

Within the parameter space allowed by the EWPO constraint, the $5\sigma$
discovery reach of Signal~2 extends to VLL masses of approximately
$2.5$, $2.8$, and $2.9$~TeV for
$\mathcal{L}=1$, $3$, and $5~\mathrm{ab}^{-1}$, respectively.
The corresponding reaches for Signal~3 are approximately
$2.4$, $2.7$, and $2.8$~TeV.

A complementary boosted-object analysis of the charged-VLL channels is
presented in Appendix~\ref{app:fatjet}. The resulting sensitivities are
comparable to those obtained with the baseline resolved-jet reconstruction,
providing an additional validation of the search strategy in the high-mass
region.
\section{SUMMARY AND CONCLUSIONS}
\label{sec:conclusion}

Future high-energy $e^+e^-$ colliders provide a clean environment for
searching for new EW fermions. In this work, the sensitivity of
the 3~TeV CLIC to first-generation
$\mathrm{SU}(2)_L$ doublet VLLs has been investigated through
single-production processes. Three complementary final states have been
studied:
$e^\pm+2j+\slashed{E}_T$ from the neutral-VLL channel,
and $e^+e^-+2b$ and $e^+e^-+2j$ from the charged-VLL decay modes
$E^\pm\to e^\pm h$ and $E^\pm\to e^\pm Z$, respectively.
Signal and background samples have been simulated including ISR, beamstrahlung, and detector effects, and cut-based analyses have
been performed for integrated luminosities of
$\mathcal{L}=1$, $3$, and $5~\mathrm{ab}^{-1}$.

The main findings are summarized as follows.

\begin{itemize}

\item
The neutral-VLL production channel
$e^+e^-\to\nu_e\bar N+\bar\nu_eN$
provides the highest sensitivity among the channels considered, benefiting
from its larger production rate and favorable signal topology. For
$\mathcal{L}=1~\mathrm{ab}^{-1}$, the expected $2\sigma$ exclusion
sensitivity reaches
$\varepsilon\simeq0.013$ at $M_N=1.3$~TeV and
$\varepsilon\simeq0.065$ at $M_N=2.9$~TeV.
The corresponding $5\sigma$ discovery sensitivity extends from
approximately $\varepsilon\simeq0.02$ to $0.10$.

\item
The charged-VLL channels provide complementary sensitivity through the
$E^\pm\to e^\pm h$ and $E^\pm\to e^\pm Z$ decay modes. At
$\mathcal{L}=5~\mathrm{ab}^{-1}$, the expected $2\sigma$ exclusion reaches
are approximately
$\varepsilon\simeq0.06$--$0.22$ and
$\varepsilon\simeq0.06$--$0.26$
for the Higgs and $Z$ channels, respectively.
Within the parameter space allowed by the EWPOs, the
corresponding $5\sigma$ discovery reaches extend to VLL masses of
approximately $2.9$~TeV and $2.8$~TeV, respectively.

\item
The comparison with EWPOs highlights the
complementarity between direct collider searches and indirect measurements.
The neutral-VLL channel probes mixing parameters well below the current
EWPO limit over the considered mass range, while the charged-VLL channels
remain sensitive to regions approaching the indirect constraints at lower
masses and become limited by phase-space suppression near the kinematic
boundary.

\end{itemize}

A complementary boosted-object analysis of the charged-VLL channels,
presented in Appendix~\ref{app:fatjet}, yields sensitivities comparable to
those obtained with the baseline resolved-jet reconstruction, providing an
additional validation of the search strategy in the multi-TeV regime.
Since the charged and neutral components of the VLL doublet are nearly
degenerate in mass, a statistical combination of the three channels may
further improve the overall sensitivity. A complete treatment including
correlated experimental systematic uncertainties and channel combinations
would require a dedicated experimental analysis.

Overall, the results demonstrate that single production at a 3~TeV CLIC
offers a sensitive probe of first-generation doublet VLLs,
particularly in the small-mixing regime. The complementary neutral and
charged channels provide direct access to the VLL--SM lepton mixing
interaction and extend the sensitivity to VLL masses approaching the
kinematic limit of the collider.
\begin{acknowledgments}
\vspace*{-0.5truecm}
Y.-B. L.  is supported by the Natural Science Foundation of Henan Province (Grant No. 252300421988).
S. M. is supported  through the NExT Institute and STFC CG ST/X000583/1.
\end{acknowledgments}
\appendix

\section{Fat-jet reconstruction for the charged channels}
\label{app:fatjet}

For the VLL mass range considered in this work,
$M_L=1.3$--$2.9$~TeV, the EW bosons from
$E^\pm$ decays can become highly boosted, causing their hadronic decay
products to be increasingly collimated. Therefore, the resolved-jet
reconstruction adopted in the baseline analysis may suffer from reduced
efficiency in the high-mass region.
To assess the robustness of our results and the possible benefit of
boosted-boson reconstruction, we perform a complementary fat-jet analysis for
the two charged-VLL channels, Signal~2 and Signal~3.

As a complementary approach, the hadronic Higgs and $Z$ decays are
reconstructed as single fat-jet candidates using the VLC algorithm with an
enlarged radius parameter $R=1.0$. The fat jet is required to satisfy
$p_T^J>400$~GeV, together with
$|M_J-M_h|<20$~GeV for Signal~2 or
$|M_J-M_Z|<15$~GeV for Signal~3. The resolved dijet observables in the
baseline analysis are correspondingly replaced by the fat-jet observables
$M_J$, $M_{e_1J}$, and $\Delta R_{e_1J}$, while the electron-selection
requirements are kept unchanged.

\subsection{Signal 2 with fat-jet reconstruction}

\begin{figure}[htbp]
\begin{center}
\centerline{\hspace{3.0cm}\epsfxsize=10cm\epsffile{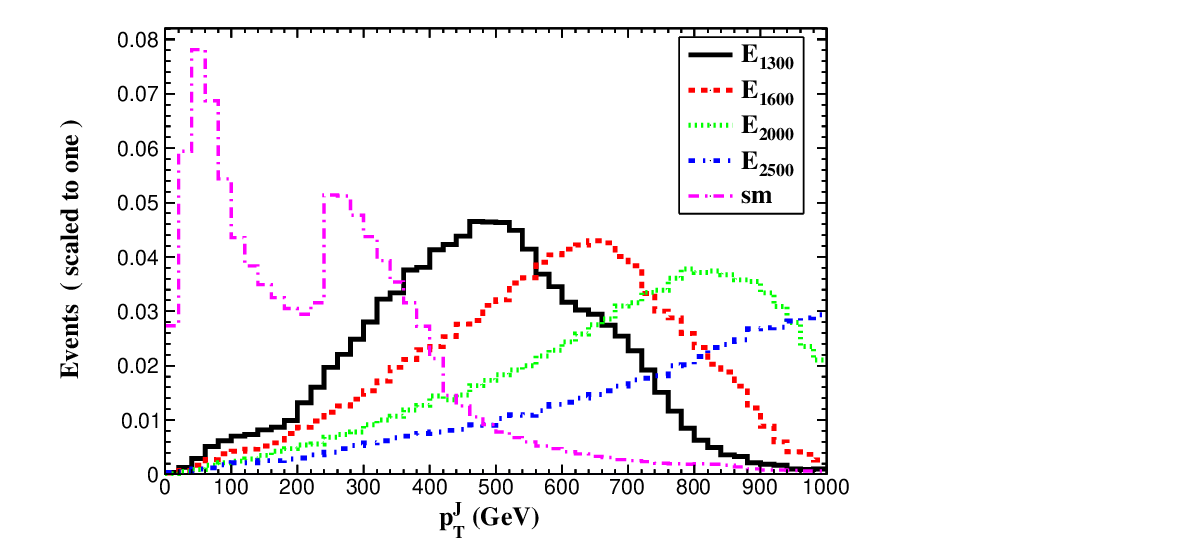}
\hspace{-3.0cm}\epsfxsize=10cm\epsffile{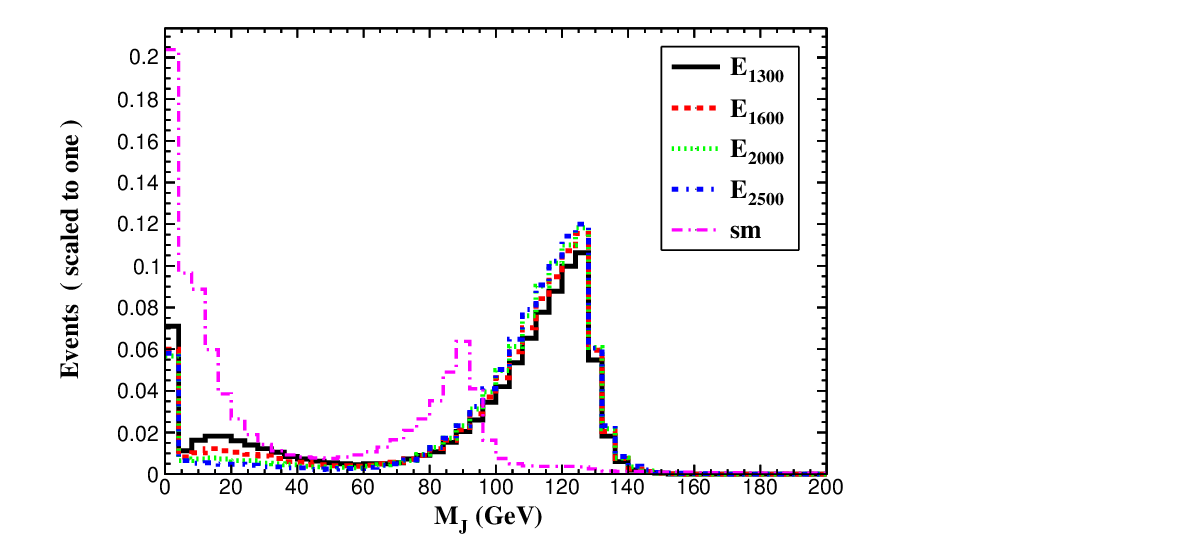}}
\caption{Normalized distributions of (a) the fat-jet transverse momentum
$p_T^J$ and (b) the fat-jet invariant mass $M_J$ for Signal~2 and the dominant
SM background at $\sqrt{s}=3$~TeV.}
\label{fig:hbb-fatjet}
\end{center}
\end{figure}

For Signal~2,
$E^\pm\to e^\pm h\to e^\pm b\bar b$,
the Higgs-boson candidate is reconstructed as a single fat jet with $R=1.0$.
Figure~\ref{fig:hbb-fatjet} shows the normalized distributions of the fat-jet
transverse momentum $p_T^J$ and invariant mass $M_J$ for representative
signal benchmarks and the dominant SM background. The signal exhibits a
harder $p_T^J$ spectrum and a resonance structure around the Higgs-boson mass,
allowing additional separation from the continuum background.

The event selection follows the baseline strategy, with the resolved
Higgs-boson reconstruction replaced by the fat-jet candidate. The following
requirements are imposed:
\begin{itemize}

\item[(i)]
Exactly two isolated, oppositely charged electrons satisfying
\[
p_T^{e_1}>500~\mathrm{GeV},
\qquad
p_T^{e_2}>10~\mathrm{GeV}.
\]

\item[(ii)]
At least one fat jet satisfying
\[
p_T^J>400~\mathrm{GeV},
\qquad
|M_J-M_h|<20~\mathrm{GeV}.
\]

\item[(iii)]
The reconstructed heavy-VLL candidate is required to satisfy
\[
M_{e_1J}>1200~\mathrm{GeV},
\qquad
2.5<\Delta R_{e_1J}<3.5.
\]
The latter requirement exploits the nearly back-to-back topology inherited
from the two-body decay $E^\pm\to e^\pm h$.

\end{itemize}

\begin{table}[htbp]
\centering
\caption{Cutflow for Signal~2 with fat-jet reconstruction using the VLC
algorithm with $R=1.0$. The entries denote cross sections in fb for the
$e^+e^-+J$ final state with $\varepsilon=0.1$. The numbers in parentheses
give the corresponding results of the baseline resolved-jet analysis
($N=2$, $R=0.5$).}
\label{cut:hbb-fatjet}
\begin{tabular}{c|cccc|c}
\hline
Cuts & \multicolumn{4}{c|}{Signal ($M_E$ in GeV)} & Background \\
\cline{2-5}
& 1300 & 1600 & 2000 & 2500 & $e^+e^-jj$ \\
\hline
Basic selection
& 0.084 (0.084) & 0.041 (0.041) & 0.023 (0.023) & 0.011 (0.011)
& 14.8 (14.8) \\
Full selection
& 0.018 (0.010) & 0.015 (0.0086) & 0.009 (0.0059) & 0.0042 (0.0028)
& 0.035 (0.012) \\
\hline
\end{tabular}
\end{table}

The corresponding cutflow is summarized in Table~\ref{cut:hbb-fatjet}.
The fat-jet reconstruction retains a larger fraction of the signal than the
resolved analysis, with the final signal cross section ranging from
$0.018$~fb at $M_E=1300$~GeV to $0.0042$~fb at $M_E=2500$~GeV. The surviving
background cross section is $0.035$~fb, compared with $0.012$~fb in the
resolved analysis. The fat-jet reconstruction increases the signal acceptance but also retains a
larger background contribution. As a result, the improvement in signal yield
is partially compensated by the increased background level, leading to a
sensitivity comparable to the baseline resolved-jet strategy.

\subsection{Signal 3 with fat-jet reconstruction}

For Signal~3,
$E^\pm\to e^\pm Z\to e^\pm q\bar q$,
the hadronic $Z$-boson decay is similarly reconstructed as a single fat jet
with $R=1.0$. The event selection requires two isolated, oppositely charged
electrons satisfying
\[
p_T^{e_1}>500~\mathrm{GeV},
\qquad
p_T^{e_2}>10~\mathrm{GeV},
\]
together with a fat-jet candidate satisfying
\[
p_T^J>400~\mathrm{GeV},
\qquad
|M_J-M_Z|<15~\mathrm{GeV}.
\]
The heavy-VLL reconstruction further requires
\[
M_{e_1J}>1200~\mathrm{GeV},
\qquad
2.5<\Delta R_{e_1J}<3.5,
\]
where the angular requirement again exploits the nearly back-to-back topology
associated with the two-body decay $E^\pm\to e^\pm Z$.

\begin{table}[htbp]
\centering
\caption{Cutflow for Signal~3 with fat-jet reconstruction using the VLC
algorithm with $R=1.0$. The entries denote cross sections in fb for the
$e^+e^-+J$ final state with $\varepsilon=0.1$. The numbers in parentheses
give the corresponding results of the baseline resolved-jet analysis
($N=2$, $R=0.5$).}
\label{cut:zqq-fatjet}
\begin{tabular}{c|cccc|c}
\hline
Cuts & \multicolumn{4}{c|}{Signal ($M_E$ in GeV)} & Background \\
\cline{2-5}
& 1300 & 1600 & 2000 & 2500 & $e^+e^-jj$ \\
\hline
Basic selection
& 0.160 (0.160) & 0.096 (0.096) & 0.054 (0.054) & 0.025 (0.025)
& 14.8 (14.8) \\
Full selection
& 0.043 (0.042) & 0.039 (0.036) & 0.024 (0.021) & 0.011 (0.0094)
& 0.20 (0.255) \\
\hline
\end{tabular}
\end{table}

The corresponding results are summarized in Table~\ref{cut:zqq-fatjet}.
After the full selection, the background cross section is reduced to
$0.20$~fb, compared with $0.255$~fb in the resolved analysis, while the signal
cross sections range from $0.043$~fb at $M_E=1300$~GeV to $0.011$~fb at
$M_E=2500$~GeV. In contrast to the Higgs channel, the fat-jet strategy
slightly improves both the signal acceptance and the background rejection in
the hadronic $Z$ channel.

\subsection{Sensitivity comparison}

\begin{figure}[htbp]
\begin{center}
\centerline{\hspace{1.0cm}\epsfxsize=9cm\epsffile{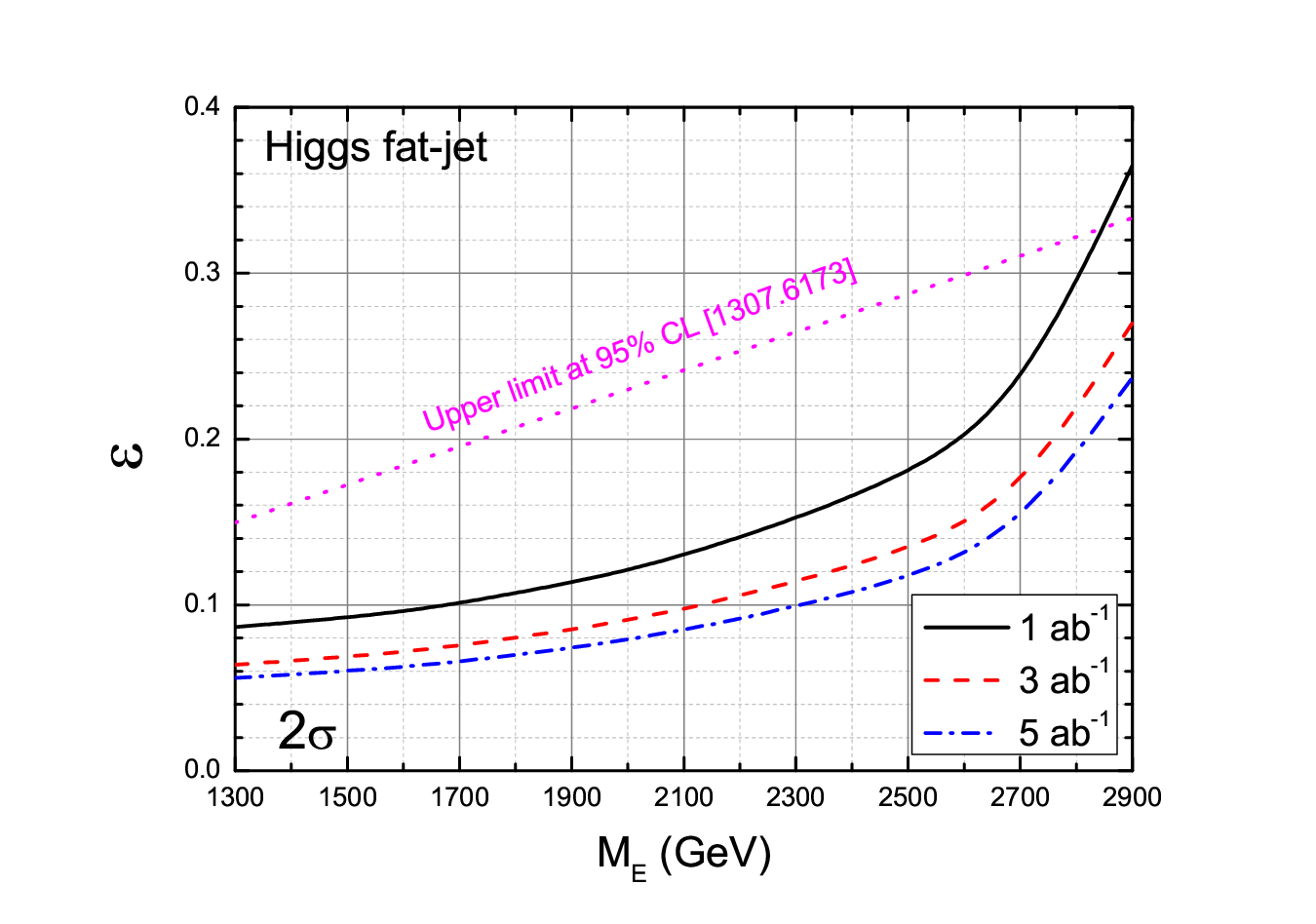}
\hspace{-1.0cm}\epsfxsize=9cm\epsffile{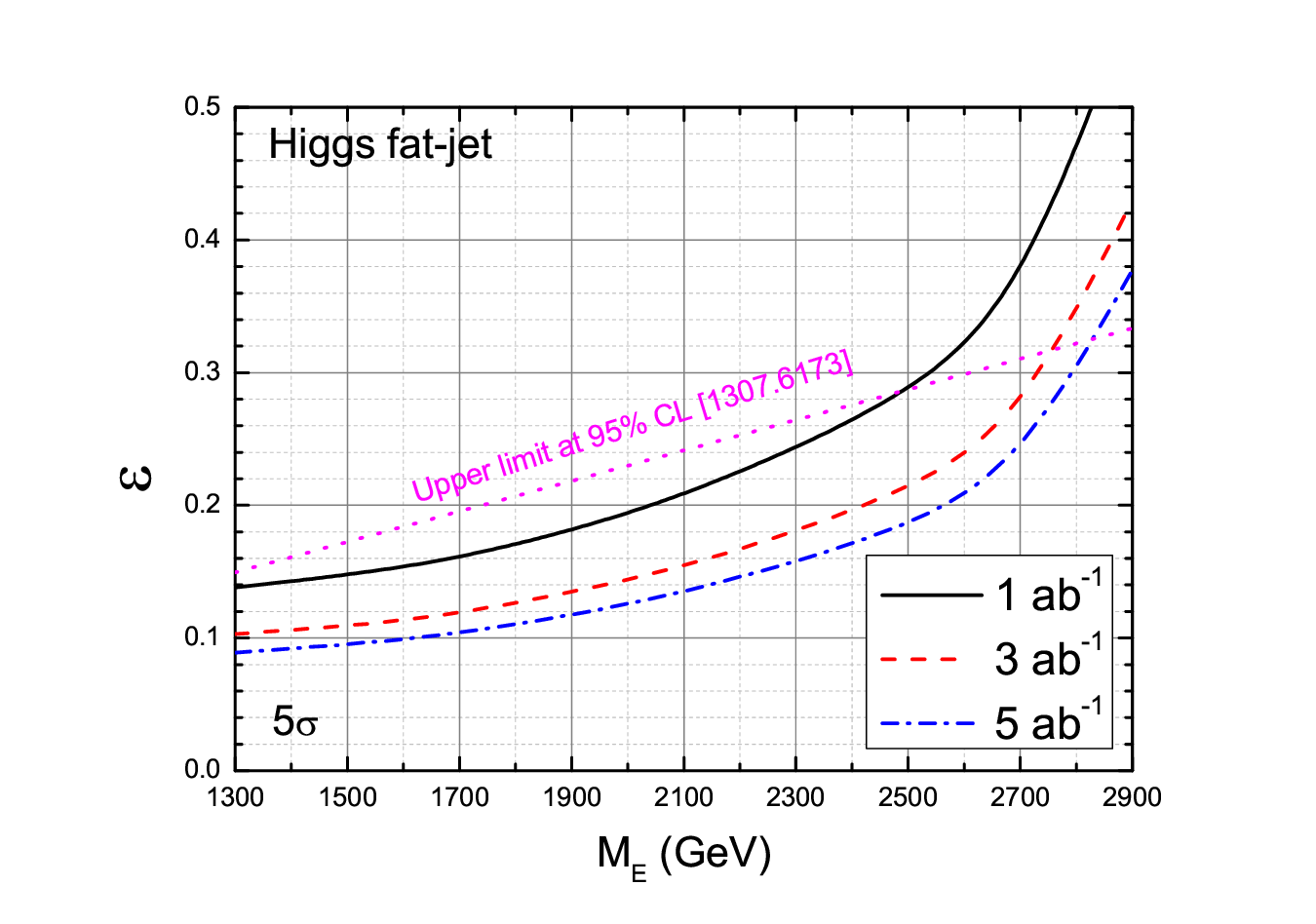}}
\centerline{\hspace{1.0cm}\epsfxsize=9cm\epsffile{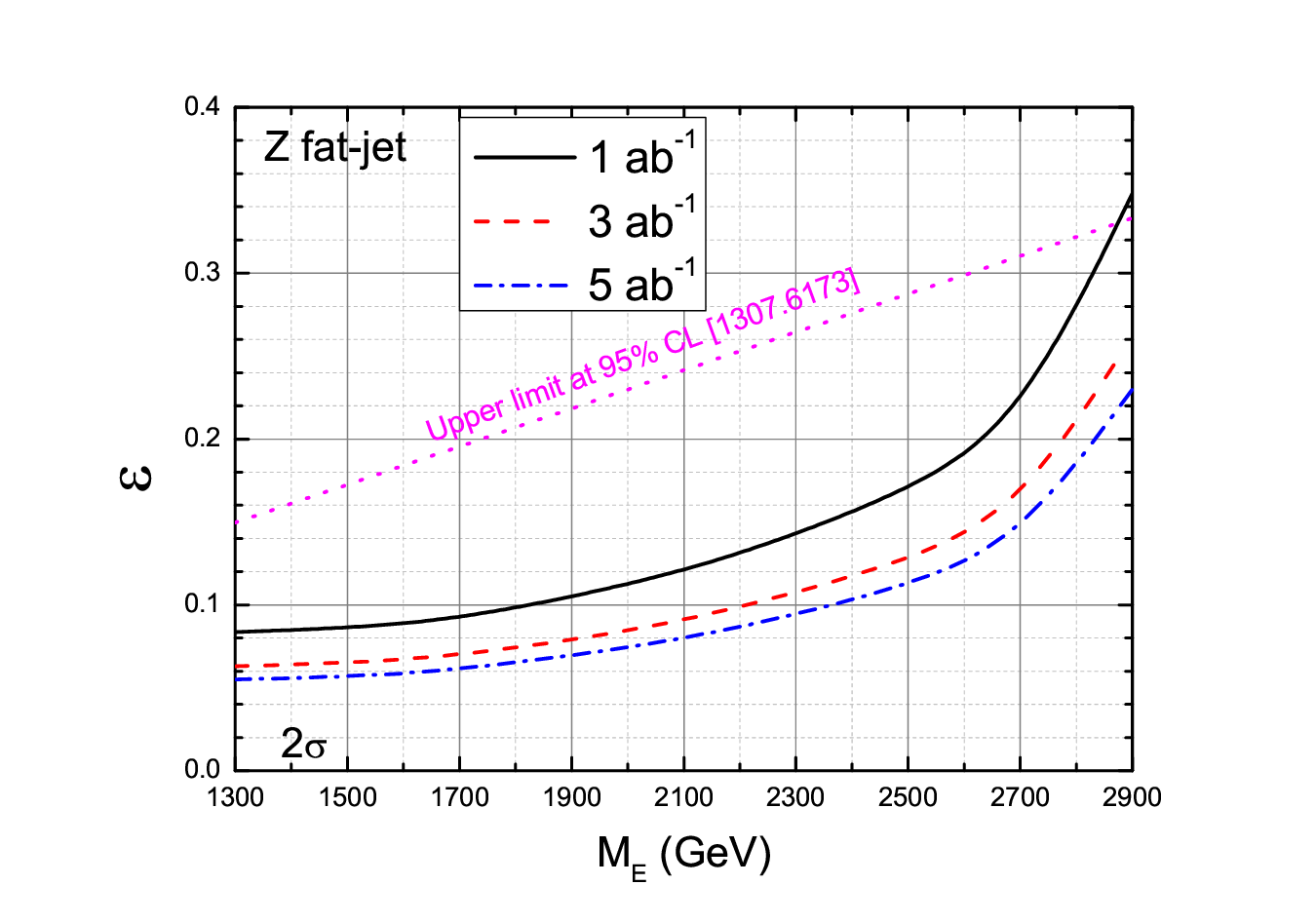}
\hspace{-1.0cm}\epsfxsize=9cm\epsffile{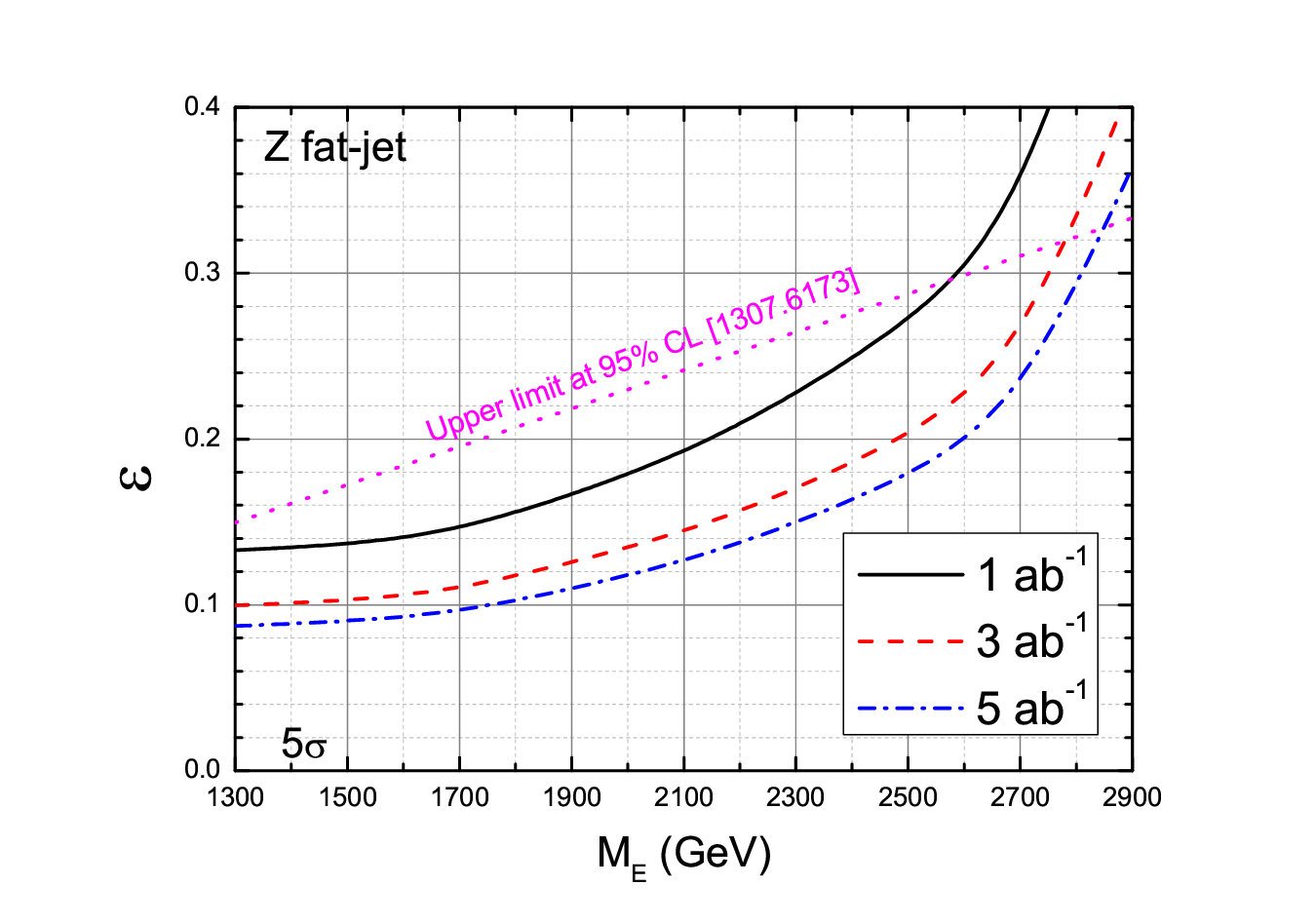}}
\caption{Expected $2\sigma$ exclusion (left) and $5\sigma$ discovery (right)
reaches for Signal~2 (top) and Signal~3 (bottom) obtained with the fat-jet
reconstruction at $\sqrt{s}=3$~TeV for
$\mathcal{L}=1$, $3$, and $5~\mathrm{ab}^{-1}$.}
\label{fig-sensitivity-fat}
\end{center}
\end{figure}

\begin{figure}[htbp]
\begin{center}
\centerline{\hspace{1.0cm}\epsfxsize=9cm\epsffile{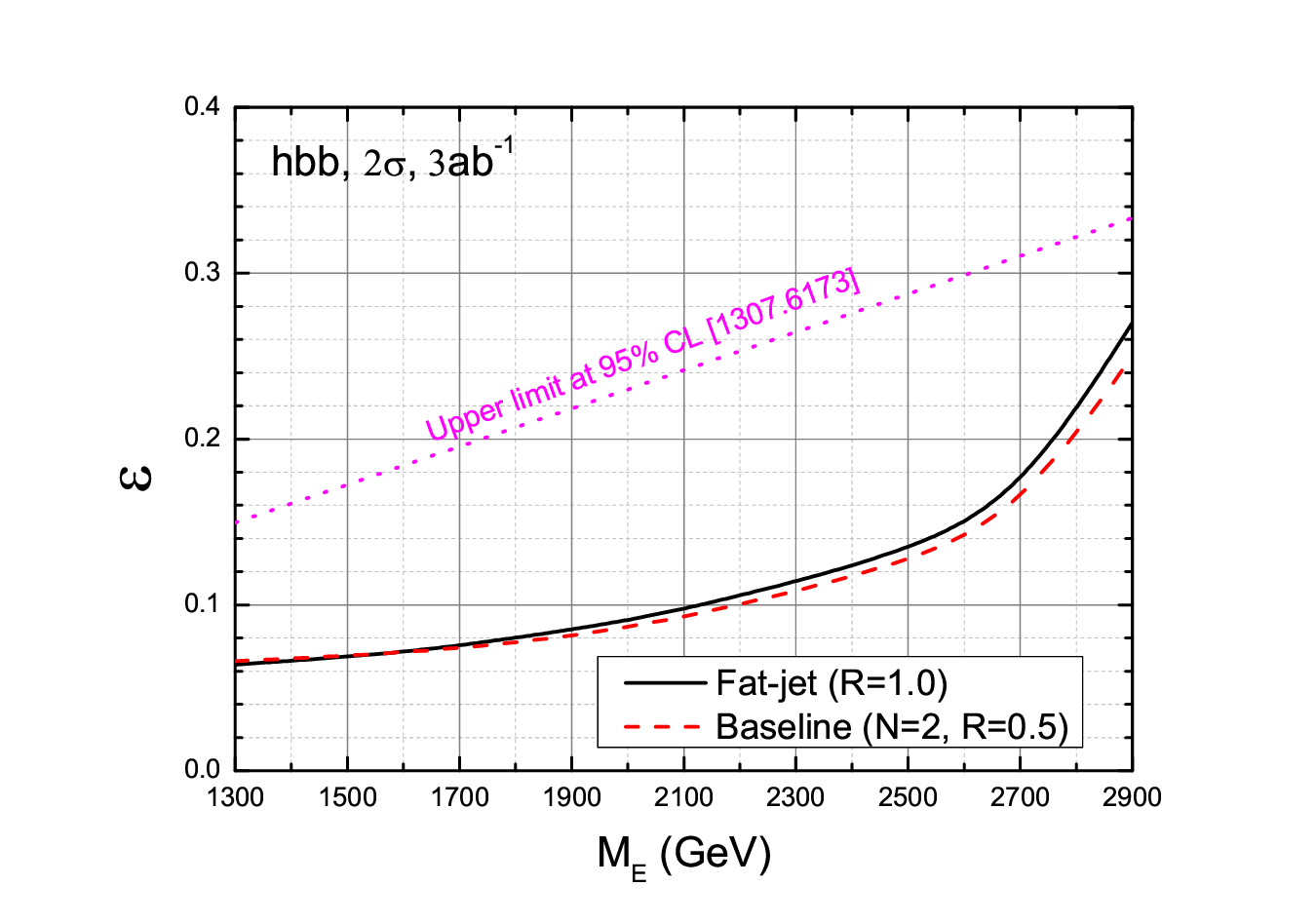}
\hspace{-1.0cm}\epsfxsize=9cm\epsffile{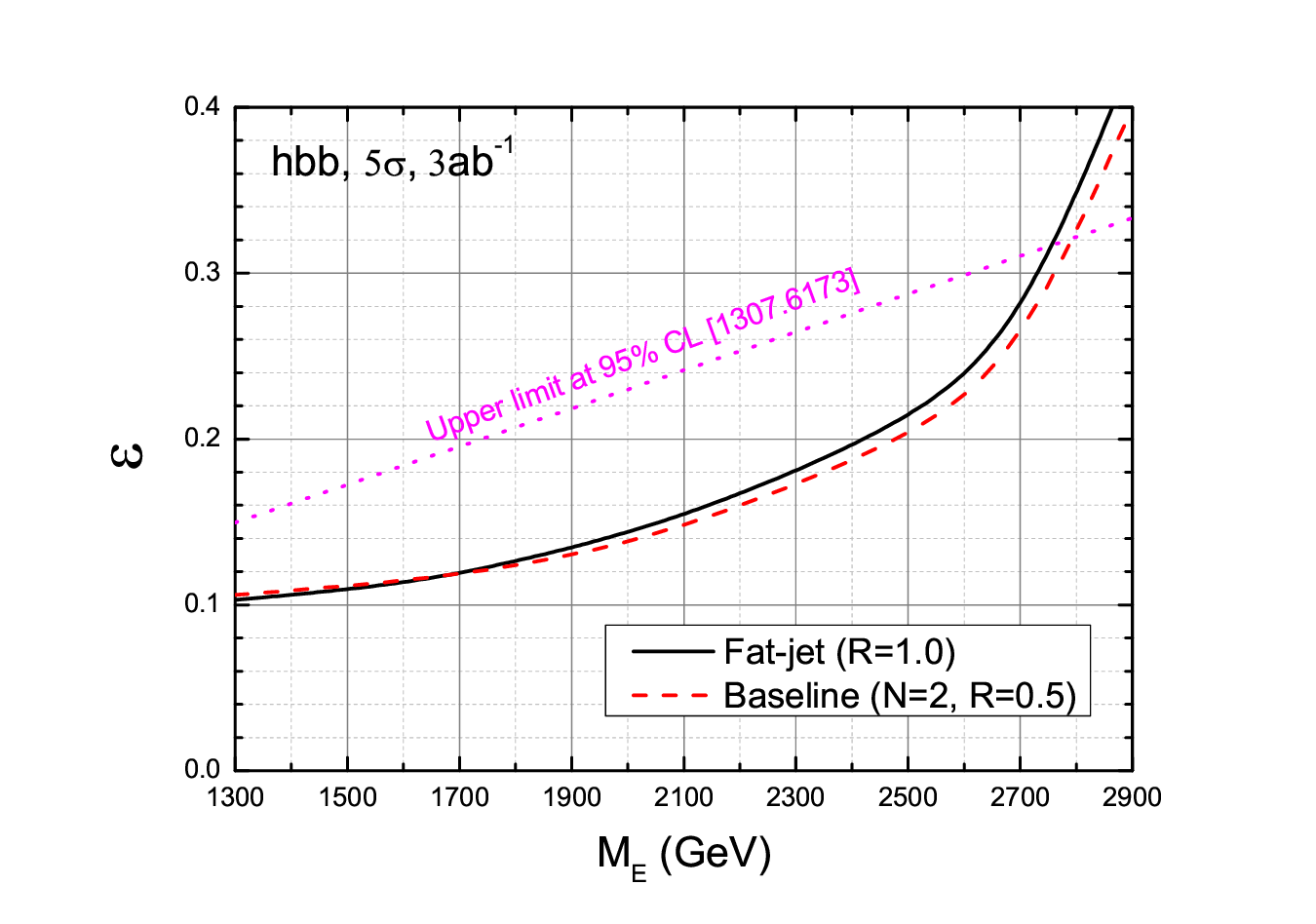}}
\centerline{\hspace{1.0cm}\epsfxsize=9cm\epsffile{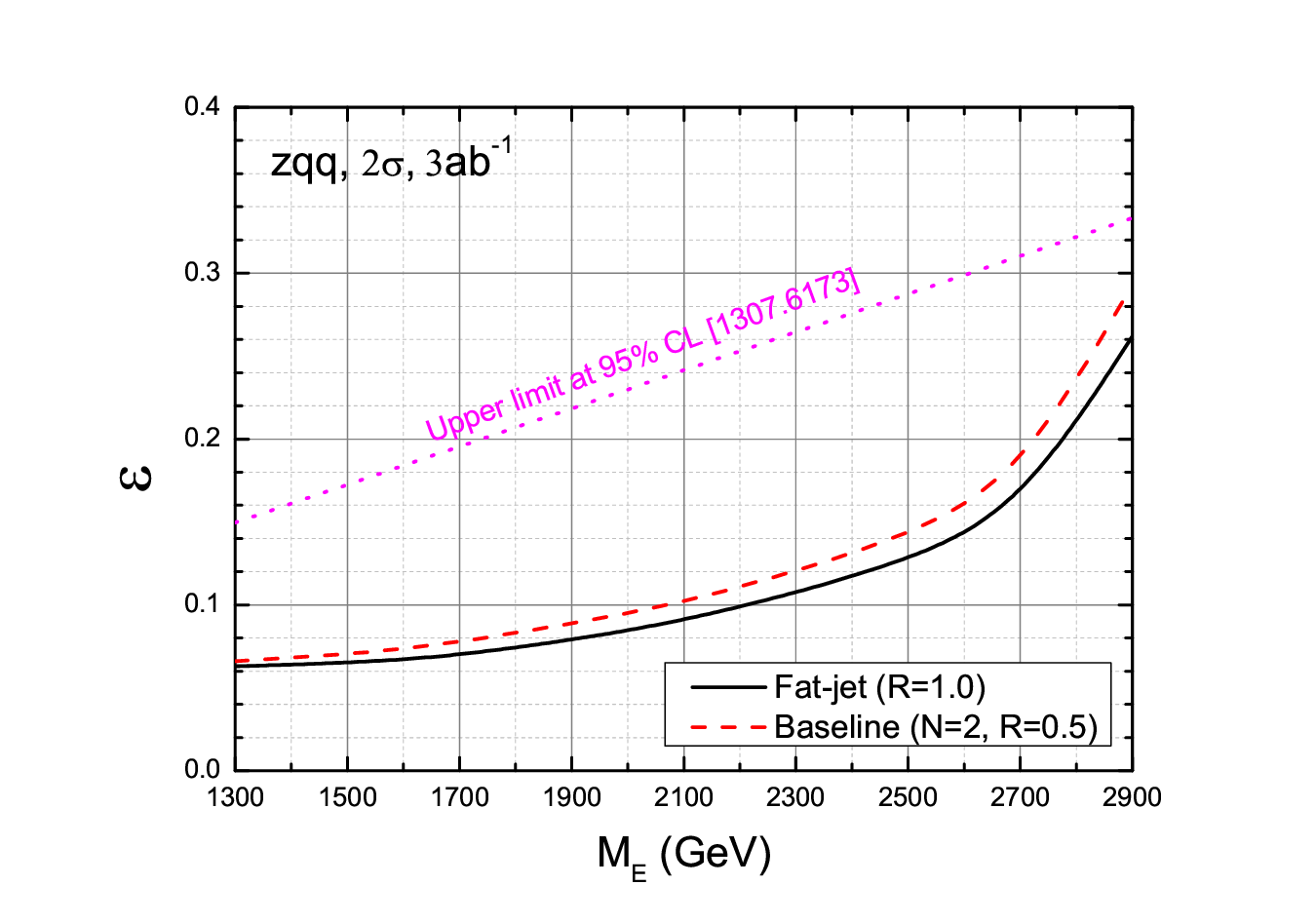}
\hspace{-1.0cm}\epsfxsize=9cm\epsffile{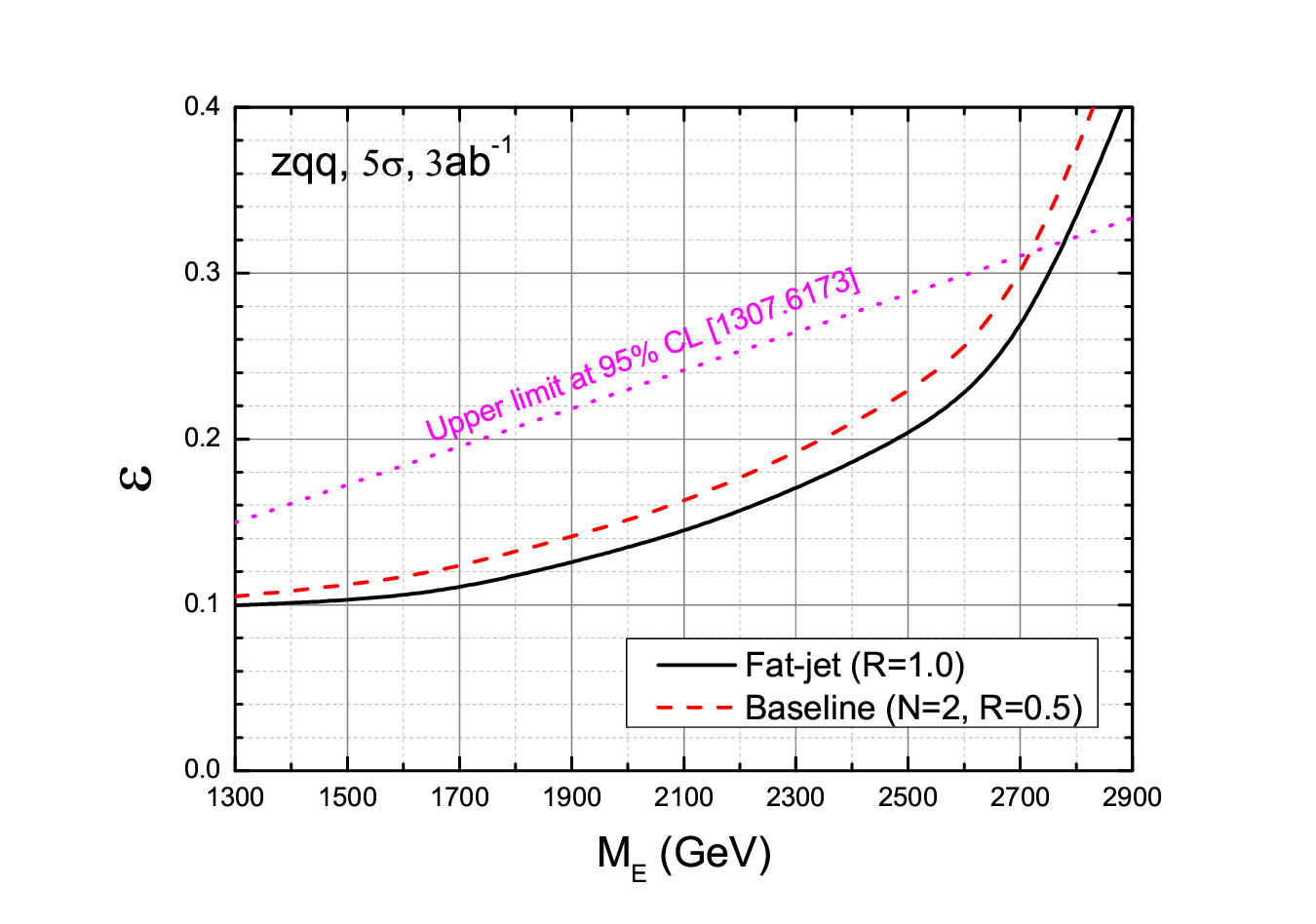}}
\caption{Comparison of the $2\sigma$ exclusion (left) and $5\sigma$
discovery (right) reaches for Signal~2 (top, $h\to b\bar b$) and Signal~3
(bottom, $Z\to q\bar q$) at $\sqrt{s}=3$~TeV and
$\mathcal{L}=3~\mathrm{ab}^{-1}$. The solid and dashed curves correspond to
the fat-jet reconstruction (VLC with $R=1.0$) and the baseline resolved-jet
analysis (VLC with $N=2$ and $R=0.5$), respectively. The dotted curve denotes
the EWPO upper bound~\cite{deBlas:2013gla}.}
\label{fig:fatjet-comparison}
\end{center}
\end{figure}

Figure~\ref{fig-sensitivity-fat} shows the exclusion and discovery reaches
obtained with the fat-jet reconstruction for Signal~2 and Signal~3 at
integrated luminosities of $1$, $3$, and $5~\mathrm{ab}^{-1}$. A direct
comparison with the baseline resolved-jet analysis at
$\mathcal{L}=3~\mathrm{ab}^{-1}$ is presented in
Fig.~\ref{fig:fatjet-comparison}. Overall, the two reconstruction strategies
provide comparable sensitivities over the mass range considered.

For Signal~2, the increased signal acceptance obtained with the fat-jet
reconstruction is offset by the larger residual background, resulting in a
slightly weaker or comparable sensitivity. For Signal~3, the fat-jet
reconstruction provides a modest improvement, particularly in the
intermediate-mass region. For example, at $M_E=2000$~GeV and
$\mathcal{L}=3~\mathrm{ab}^{-1}$, the $2\sigma$ exclusion reach changes from
$\varepsilon=0.086$ to $0.090$ for Signal~2 and from
$\varepsilon=0.094$ to $0.084$ for Signal~3 when switching from the resolved-
to the fat-jet reconstruction. The differences between the two approaches
remain at the $\mathcal{O}(10\%)$ level over the mass range studied.

Given their comparable overall performance, we retain the resolved-jet
strategy as the baseline approach in the main analysis and use the fat-jet
results as a complementary validation of the search strategy in the boosted
regime.


\begin{thebibliography}{99}
\bibitem{Huang:2012kz}
G.~Y.~Huang, K.~Kong and S.~C.~Park,
\href{doi:10.1007/JHEP06(2012)099}{JHEP \textbf{06}, 099 (2012)}.
[arXiv:1204.0522 [hep-ph]].

\bibitem{Kong:2010qd}
K.~Kong, S.~C.~Park and T.~G.~Rizzo,
\href{doi:10.1007/JHEP07(2010)059}{JHEP \textbf{07}, 059 (2010)}.
[arXiv:1004.4635 [hep-ph]].

\bibitem{Graham:2009gy}
P.~W.~Graham, A.~Ismail, S.~Rajendran and P.~Saraswat,
\href{doi:10.1103/PhysRevD.81.055016}{Phys.\ Rev.\ D \textbf{81}, 055016 (2010)}.
[arXiv:0910.3020 [hep-ph]].

\bibitem{Endo:2011xq}
M.~Endo, K.~Hamaguchi, S.~Iwamoto and N.~Yokozaki,
\href{doi:10.1103/PhysRevD.85.095012}{Phys.\ Rev.\ D \textbf{85}, 095012 (2012)}.
[arXiv:1112.5653 [hep-ph]].

\bibitem{Martin:2012dg}
S.~P.~Martin and J.~D.~Wells,
\href{doi:10.1103/PhysRevD.86.035017}{Phys.\ Rev.\ D \textbf{86}, 035017 (2012)}.
[arXiv:1206.2956 [hep-ph]].

\bibitem{Endo:2012cc}
M.~Endo, K.~Hamaguchi, K.~Ishikawa, S.~Iwamoto and N.~Yokozaki,
\href{doi:10.1007/JHEP01(2013)181}{JHEP \textbf{01}, 181 (2013)}.
[arXiv:1212.3935 [hep-ph]].

\bibitem{Fischler:2013tva}
W.~Fischler and W.~Tangarife,
\href{doi:10.1007/JHEP05(2014)151}{JHEP \textbf{05}, 151 (2014)}.
[arXiv:1310.6369 [hep-ph]].

\bibitem{DeCurtis:2018iqd}
S.~De Curtis, L.~Delle Rose, S.~Moretti and K.~Yagyu,
\href{doi:10.1016/j.physletb.2018.09.042}{Phys. Lett. B \textbf{786}, 189 (2018)}.
[arXiv:1803.01865 [hep-ph]].

\bibitem{DeCurtis:2018zvh}
S.~De Curtis, L.~Delle Rose, S.~Moretti and K.~Yagyu,
\href{doi:10.1007/JHEP12(2018)051}{JHEP \textbf{12}, 051 (2018)}.
[arXiv:1810.06465 [hep-ph]].

\bibitem{He:1999vp}
H.~J.~He, T.~M.~P.~Tait and C.~P.~Yuan,
\href{doi:10.1103/PhysRevD.62.011702}{Phys.\ Rev.\ D \textbf{62}, 011702(R) (2000)}.
[arXiv:hep-ph/9911266 [hep-ph]].

\bibitem{Wang:2013jwa}
X.~F.~Wang, C.~Du and H.~J.~He,
\href{doi:10.1016/j.physletb.2013.05.015}{Phys.\ Lett.\ B \textbf{723}, 314-323 (2013)}.
[arXiv:1304.2257 [hep-ph]].

\bibitem{He:2001fz}
H.~J.~He, C.~T.~Hill and T.~M.~P.~Tait,
\href{doi:10.1103/PhysRevD.65.055006}{Phys.\ Rev.\ D \textbf{65}, 055006 (2002)}.
[arXiv:hep-ph/0108041 [hep-ph]].

\bibitem{He:2014ora}
H.~J.~He and Z.~Z.~Xianyu,
\href{doi:10.1088/1475-7516/2014/10/019}{JCAP\ \textbf{10}, 019 (2014)}.
[arXiv:1405.7331 [hep-ph]].


\bibitem{Arkani-Hamed:2012dcq}
N.~Arkani-Hamed, K.~Blum, R.~T.~D'Agnolo and J.~Fan,
\href{doi:10.1007/JHEP01(2013)149}{JHEP \textbf{01}, 149 (2013)}.
[arXiv:1207.4482 [hep-ph]].

\bibitem{Schwaller:2013hqa}
P.~Schwaller, T.~M.~P.~Tait and R.~Vega-Morales,
\href{doi:10.1103/PhysRevD.88.035001}{Phys.\ Rev.\ D \textbf{88}, no.3, 035001 (2013)}.
[arXiv:1305.1108 [hep-ph]].

\bibitem{Halverson:2014nwa}
J.~Halverson, N.~Orlofsky and A.~Pierce,
\href{doi:10.1103/PhysRevD.90.015002}{Phys.\ Rev.\ D \textbf{90}, no.1, 015002 (2014)}.
[arXiv:1403.1592 [hep-ph]].

\bibitem{Bahrami:2016has}
S.~Bahrami, M.~Frank, D.~K.~Ghosh, N.~Ghosh and I.~Saha,
\href{doi:10.1103/PhysRevD.95.095024}{Phys.\ Rev.\ D \textbf{95}, no.9, 095024 (2017)}.
[arXiv:1612.06334 [hep-ph]].

\bibitem{Bhattacharya:2018fus}
S.~Bhattacharya, P.~Ghosh, N.~Sahoo and N.~Sahu,
\href{doi:10.3389/fphy.2019.00080}{Front.\ in Phys.\ \textbf{7}, 80 (2019)}.
[arXiv:1812.06505 [hep-ph]].

\bibitem{Xiao:2014kba}
M.~L.~Xiao and J.~H.~Yu,
\href{doi:10.1103/PhysRevD.90.014007}{Phys.\ Rev.\ D \textbf{90}, no.1, 014007 (2014)}.
[arXiv:1404.0681 [hep-ph]].

\bibitem{Cingiloglu:2024vdh}
K.~Y.~Cingiloglu and M.~Frank,
\href{doi:10.1103/PhysRevD.111.016025}{Phys.\ Rev.\ D \textbf{111}, no.1, 016025 (2025)}.
[arXiv:2408.10898 [hep-ph]].

\bibitem{Hiller:2019mou}
G.~Hiller, C.~Hormigos-Feliu, D.~F.~Litim and T.~Steudtner,
\href{doi:10.1103/PhysRevD.102.071901}{Phys.\ Rev.\ D \textbf{102}, no.7, 071901 (2020)}.
[arXiv:1910.14062 [hep-ph]].

\bibitem{DeJesus:2020yqx}
A.~S.~De Jesus, S.~Kovalenko, F.~S.~Queiroz, C.~Siqueira and K.~Sinha,
\href{doi:10.1103/PhysRevD.102.035004}{Phys.\ Rev.\ D \textbf{102}, no.3, 035004 (2020)}.
[arXiv:2004.01200 [hep-ph]].

\bibitem{Frank:2020smf}
M.~Frank and I.~Saha,
\href{doi:10.1103/PhysRevD.102.115034}{Phys.\ Rev.\ D \textbf{102}, no.11, 115034 (2020)}.
[arXiv:2008.11909 [hep-ph]].

\bibitem{Dermisek:2021ajd}
R.~Dermisek, K.~Hermanek and N.~McGinnis,
\href{doi:10.1103/PhysRevD.104.055033}{Phys.\ Rev.\ D \textbf{104}, no.5, 055033 (2021)}.
[arXiv:2103.05645 [hep-ph]].

\bibitem{Brune:2022rlo}
T.~Brune, T.~W.~Kephart and H.~P\"as,
\href{doi:10.1140/epjc/s10052-024-13617-5}{Eur.\ Phys.\ J.\ C \textbf{84}, no.12, 1254 (2024)}.
[arXiv:2205.05566 [hep-ph]].

\bibitem{Guedes:2022cfy}
G.~Guedes and P.~Olgoso,
\href{doi:10.1007/JHEP09(2022)181}{JHEP \textbf{09}, 181 (2022)}.
[arXiv:2205.04480 [hep-ph]].

\bibitem{Erdelyi:2025axy}
B.~A.~Erdelyi, R.~Gr{\"o}ber and N.~Selimovic,
\href{doi:10.1007/JHEP05(2025)135}{JHEP \textbf{05} (2025), 135}.
[arXiv:2501.07628 [hep-ph]].
\bibitem{Athron:2025ets}
P.~Athron, K.~M{\"o}hling, D.~St{\"o}ckinger and H.~St{\"o}ckinger-Kim,
\href{doi:10.1016/j.ppnp.2025.104225}{Prog. Part. Nucl. Phys. \textbf{148} (2026), 104225}.
[arXiv:2507.09289 [hep-ph]].


\bibitem{He:2022zjz}
S.~P.~He,
\href{doi:10.1088/1674-1137/ac9e4c}{Chin.\ Phys.\ C \textbf{47}, no.4, 043102 (2023)}.
[arXiv:2205.02088 [hep-ph]].

\bibitem{Kawamura:2022fhm}
J.~Kawamura and S.~Raby,
\href{doi:10.1103/PhysRevD.106.035009}{Phys.\ Rev.\ D \textbf{106}, no.3, 035009 (2022)}.
[arXiv:2205.10480 [hep-ph]].

\bibitem{delAguila:2008pw}
F.~del Aguila, J.~de Blas and M.~Perez-Victoria,
\href{doi:10.1103/PhysRevD.78.013010}{Phys.\ Rev.\ D \textbf{78}, 013010 (2008)}.
[arXiv:0803.4008 [hep-ph]].

\bibitem{Ishiwata:2013gma}
K.~Ishiwata and M.~B.~Wise,
\href{doi:10.1103/PhysRevD.88.055009}{Phys.\ Rev.\ D \textbf{88}, 055009 (2013)}.
[arXiv:1307.1112 [hep-ph]].

\bibitem{Kearney:2012zi}
J.~Kearney, A.~Pierce and N.~Weiner,
\href{doi:10.1103/PhysRevD.86.113005}{Phys.\ Rev.\ D \textbf{86}, 113005 (2012)}.
[arXiv:1207.7062 [hep-ph]].

\bibitem{Altmannshofer:2013zba}
W.~Altmannshofer, M.~Bauer and M.~Carena,
\href{doi:10.1007/JHEP01(2014)060}{JHEP \textbf{01}, 060 (2014)}.
[arXiv:1308.1987 [hep-ph]].

\bibitem{Falkowski:2013jya}
A.~Falkowski, D.~M.~Straub and A.~Vicente,
\href{doi:10.1007/JHEP05(2014)092}{JHEP \textbf{05}, 092 (2014)}.
[arXiv:1312.5329 [hep-ph]].

\bibitem{Ellis:2014dza}
S.~A.~R.~Ellis, R.~M.~Godbole, S.~Gopalakrishna and J.~D.~Wells,
\href{doi:10.1007/JHEP09(2014)130}{JHEP \textbf{09}, 130 (2014)}.
[arXiv:1404.4398 [hep-ph]].

\bibitem{Ishiwata:2015cga}
K.~Ishiwata, Z.~Ligeti and M.~B.~Wise,
\href{doi:10.1007/JHEP10(2015)027}{JHEP \textbf{10}, 027 (2015)}.
[arXiv:1506.03484 [hep-ph]].

\bibitem{Dermisek:2014cia}
R.~Dermisek, A.~Raval and S.~Shin,
\href{doi:10.1103/PhysRevD.90.034023}{Phys.\ Rev.\ D \textbf{90}, 034023 (2014)}.
[arXiv:1406.7018 [hep-ph]].

\bibitem{Crivellin:2020ebi}
A.~Crivellin, F.~Kirk, C.~A.~Manzari and M.~Montull,
\href{doi:10.1007/JHEP12(2020)166}{JHEP \textbf{12}, 166 (2020)}.
[arXiv:2008.01113 [hep-ph]].

\bibitem{Endo:2020tkb}
M.~Endo and S.~Mishima,
\href{doi:10.1007/JHEP08(2020)004}{JHEP \textbf{08}, 004 (2020)}.
[arXiv:2005.03933 [hep-ph]].

\bibitem{Chakrabarty:2020jro}
N.~Chakrabarty,
\href{doi:10.1140/epjp/s13360-021-02168-3}{Eur.\ Phys.\ J.\ Plus \textbf{136}, 1183 (2021)}.
[arXiv:2010.05215 [hep-ph]].

\bibitem{Guedes:2021oqx}
G.~Guedes and J.~Santiago,
\href{doi:10.1007/JHEP01(2022)111}{JHEP \textbf{01}, 111 (2022)}.
[arXiv:2107.03429 [hep-ph]].

\bibitem{Raju:2022zlv}
M.~Raju, A.~Mukherjee and J.~P.~Saha,
\href{doi:10.1140/epjc/s10052-023-11595-8}{Eur.\ Phys.\ J.\ C \textbf{83}, 429 (2023)}.
[arXiv:2207.02825 [hep-ph]].

\bibitem{Li:2023mrw}
X.~Q.~Li, Z.~J.~Xie, Y.~D.~Yang and X.~B.~Yuan,
\href{doi:10.1016/j.nuclphysb.2024.116646}{Nucl.\ Phys.\ B \textbf{1006}, 116646 (2024)}.
[arXiv:2307.05290 [hep-ph]].

\bibitem{Dermisek:2014qca}
R.~Dermisek, J.~P.~Hall, E.~Lunghi and S.~Shin,
\href{doi:10.1007/JHEP12(2014)013}{JHEP \textbf{12}, 013 (2014)}.
[arXiv:1408.3123 [hep-ph]].

\bibitem{Dermisek:2015oja}
R.~Dermisek, E.~Lunghi and S.~Shin,
\href{doi:10.1007/JHEP02(2016)119}{JHEP \textbf{02}, 119 (2016)}.
[arXiv:1509.04292 [hep-ph]].

\bibitem{Kumar:2015tna}
N.~Kumar and S.~P.~Martin,
\href{doi:10.1103/PhysRevD.92.115018}{Phys.\ Rev.\ D \textbf{92}, 115018 (2015)}.
[arXiv:1510.03456 [hep-ph]].

\bibitem{Bhattiprolu:2019vdu}
P.~N.~Bhattiprolu and S.~P.~Martin,
\href{doi:10.1103/PhysRevD.100.015033}{Phys.\ Rev.\ D \textbf{100}, 015033 (2019)}.
[arXiv:1905.00498 [hep-ph]].

\bibitem{Dermisek:2015hue}
R.~Dermisek, E.~Lunghi and S.~Shin,
\href{doi:10.1007/JHEP05(2016)148}{JHEP \textbf{05}, 148 (2016)}.
[arXiv:1512.07837 [hep-ph]].

\bibitem{Chen:2016lsr}
C.~H.~Chen and T.~Nomura,
\href{doi:10.1140/epjc/s10052-016-4197-3}{Eur.\ Phys.\ J.\ C \textbf{76}, 353 (2016)}.
[arXiv:1602.07519 [hep-ph]].

\bibitem{Kawamura:2019rth}
J.~Kawamura, S.~Raby and A.~Trautner,
\href{doi:10.1103/PhysRevD.100.055030}{Phys.\ Rev.\ D \textbf{100}, 055030 (2019)}.
[arXiv:1906.11297 [hep-ph]].

\bibitem{Freitas:2020ttd}
F.~F.~Freitas, J.~Gon\c{c}alves, A.~P.~Morais and R.~Pasechnik,
\href{doi:10.1007/JHEP01(2021)076}{JHEP \textbf{01}, 076 (2021)}.
[arXiv:2010.01307 [hep-ph]].

\bibitem{OsmanAcar:2021plv}
A.~Osman Acar, O.~E.~Delialioglu and S.~Sultansoy,
[arXiv:2103.08222 [hep-ph]].

\bibitem{Kawamura:2021ygg}
J.~Kawamura and S.~Raby,
\href{doi:10.1103/PhysRevD.104.035007}{Phys.\ Rev.\ D \textbf{104}, 035007 (2021)}.
[arXiv:2104.04461 [hep-ph]].

\bibitem{Bonilla:2021ize}
C.~Bonilla, A.~E.~C\'arcamo Hern\'andez, J.~Gon\c{c}alves, F.~F.~Freitas, A.~P.~Morais and R.~Pasechnik,
\href{doi:10.1007/JHEP01(2022)154}{JHEP \textbf{01}, 154 (2022)}.
[arXiv:2107.14165 [hep-ph]].

\bibitem{Baspehlivan:2022qet}
F.~Baspehlivan, B.~Dagli, O.~E.~Delialioglu and S.~Sultansoy,
[arXiv:2201.08251 [hep-ph]].

\bibitem{Cao:2023smj}
Q.~H.~Cao, J.~Guo, J.~Liu, Y.~Luo and X.~P.~Wang,
\href{doi:10.1103/PhysRevD.110.015029}{Phys.\ Rev.\ D \textbf{110}, 015029 (2024)}.
[arXiv:2311.12934 [hep-ph]].

\bibitem{Bernreuther:2023uxh}
E.~Bernreuther and B.~A.~Dobrescu,
\href{doi:10.1007/JHEP07(2023)079}{JHEP \textbf{07}, 079 (2023)}.
[arXiv:2304.08509 [hep-ph]].

\bibitem{Kawamura:2023zuo}
J.~Kawamura and S.~Shin,
\href{doi:10.1007/JHEP11(2023)025}{JHEP \textbf{11}, 025 (2023)}.
[arXiv:2308.07814 [hep-ph]].

\bibitem{Bigaran:2023ris}
I.~Bigaran, B.~A.~Dobrescu and A.~Russo,
\href{doi:10.1103/PhysRevD.109.055033}{Phys. Rev. D \textbf{109}, no.5, 055033 (2024)}.
[arXiv:2312.09189 [hep-ph]].

\bibitem{Bissmann:2020lge}
S.~Bi{\ss}mann, G.~Hiller, C.~Hormigos-Feliu and D.~F.~Litim,
\href{doi:10.1140/epjc/s10052-021-08886-3}{Eur. Phys. J. C \textbf{81} no.2, 101 (2021)}.
[arXiv:2011.12964 [hep-ph]].

\bibitem{Morais:2021ead}
A.~P.~Morais, A.~Onofre, F.~F.~Freitas, J.~Gon{\c{c}}alves, R.~Pasechnik and R.~Santos,
\href{doi:10.1140/epjc/s10052-023-11314-3}{Eur. Phys. J. C \textbf{83} no.3, 232 (2023)}.
[arXiv:2108.03926 [hep-ph]].

\bibitem{ATLAS:2015qoy}
G.~Aad \textit{et al.} [ATLAS],
\href{doi:10.1007/JHEP09(2015)108}{JHEP \textbf{09}, 108 (2015)}.
[arXiv:1506.01291 [hep-ex]].

\bibitem{ATLAS:2023sbu}
G.~Aad \textit{et al.} [ATLAS],
\href{doi:10.1007/JHEP07(2023)118}{JHEP \textbf{07}, 118 (2023)}.
[arXiv:2303.05441 [hep-ex]].

\bibitem{CMS:2019hsm}
A.~M.~Sirunyan \textit{et al.} [CMS],
\href{doi:10.1103/PhysRevD.100.052003}{Phys.\ Rev.\ D \textbf{100}, 052003 (2019)}.
[arXiv:1905.10853 [hep-ex]].

\bibitem{CMS:2022nty}
A.~Tumasyan \textit{et al.} [CMS],
\href{doi:10.1103/PhysRevD.105.112007}{Phys.\ Rev.\ D \textbf{105}, 112007 (2022)}.
[arXiv:2202.08676 [hep-ex]].

\bibitem{CMS:2022cpe}
A.~Tumasyan \textit{et al.} [CMS],
\href{doi:10.1016/j.physletb.2023.137713}{Phys.\ Lett.\ B \textbf{846}, 137713 (2023)}.
[arXiv:2208.09700 [hep-ex]].

\bibitem{CMS:2024bni}
A.~Hayrapetyan \textit{et al.} [CMS],
\href{doi:10.1016/j.physrep.2024.09.012}{Phys.\ Rep.\ \textbf{1115}, 570-677 (2025)}.
[arXiv:2405.17605 [hep-ex]].

\bibitem{ATLAS:2024mrr}
G.~Aad \textit{et al.} [ATLAS],
\href{doi:10.1007/JHEP05(2025)075}{JHEP \textbf{05}, 075  (2025)}.
[arXiv:2411.07143 [hep-ex]].





\bibitem{ILC:2013jhg}
H.~Baer \textit{et al.} [ILC],
[arXiv:1306.6352 [hep-ph]].

\bibitem{ILCInternationalDevelopmentTeam:2022izu}
A.~Aryshev \textit{et al.} [ILC International Development Team],
[arXiv:2203.07622 [physics.acc-ph]].

\bibitem{Linssen:2012hp}
L.~Linssen, A.~Miyamoto, M.~Stanitzki and H.~Weerts,
[arXiv:1202.5940 [physics.ins-det]].

\bibitem{CLICDetector:2013tfe}
H.~Abramowicz \textit{et al.} [CLIC Detector and Physics Study],
[arXiv:1307.5288 [hep-ex]].

\bibitem{Franceschini:2019zsg}
R.~Franceschini,
\href{doi:10.1142/S0217751X20410158}{Int.\ J.\ Mod.\ Phys.\ A \textbf{35}, 2041015 (2020)}.
[arXiv:1902.10125 [hep-ph]].

\bibitem{Das:2020gnt}
A.~Das, S.~Mandal and T.~Modak,
\href{doi:10.1103/PhysRevD.102.033001}{Phys. Rev. D \textbf{102}, no.3, 033001 (2020)}.
[arXiv:2005.02267 [hep-ph]].

\bibitem{Das:2020uer}
A.~Das and S.~Mandal,
\href{doi:10.1016/j.nuclphysb.2021.115374}{Nucl. Phys. B \textbf{966}, 115374 (2021)}.
[arXiv:2006.04123 [hep-ph]].

\bibitem{Yang:2021dtc}
B.~Yang, J.~Li, M.~Wang and L.~Shang,
\href{doi:10.1103/PhysRevD.104.055019}{Phys.\ Rev.\ D \textbf{104}, 055019 (2021)}.

\bibitem{Shang:2021mgn}
L.~Shang, M.~Wang, Z.~Heng and B.~Yang,
\href{doi:10.1140/epjc/s10052-021-09152-2}{Eur.\ Phys.\ J.\ C \textbf{81}, 415 (2021)}.

\bibitem{Bhattacharya:2021ltd}
S.~Bhattacharya, S.~Jahedi and J.~Wudka,
\href{doi:10.1007/JHEP05(2022)009}{JHEP \textbf{05}, 009 (2022)}.
[arXiv:2106.02846 [hep-ph]].

\bibitem{Shang:2023rfv}
L.~Shang, J.~Li, X.~Jia and B.~Yang,
\href{doi:10.1016/j.nuclphysb.2022.116071}{Nucl.\ Phys.\ B \textbf{987}, 116071 (2023)}.

\bibitem{Bhattiprolu:2023yxa}
P.~N.~Bhattiprolu, S.~P.~Martin and A.~Pierce,
\href{doi:10.1103/PhysRevD.109.035009}{Phys.\ Rev.\ D \textbf{109}, 035009 (2024)}.
[arXiv:2308.08386 [hep-ph]].

\bibitem{Yue:2024sds}
C.~X.~Yue, Y.~Q.~Wang, H.~Wang, Y.~H.~Wang and S.~Li,
\href{doi:10.1016/j.nuclphysb.2024.116482}{Nucl.\ Phys.\ B \textbf{1000}, 116482 (2024)}.
[arXiv:2402.02072 [hep-ph]].

\bibitem{Yue:2024ftz}
C.~X.~Yue, Y.~Q.~Wang, X.~C.~Sun and X.~Y.~Li,
\href{doi:10.1088/1361-6471/ad9ec8}{J. Phys. G \textbf{52}, no.2, 025003 (2025)}.
[arXiv:2412.07125 [hep-ph]].
\bibitem{Yue:2025xvk}
C.~X.~Yue, M.~S.~Y.~Wang and X.~Y.~Li,
\href{doi:10.1016/j.nuclphysb.2025.117143}{Nucl. Phys. B \textbf{1020}, 117143 (2025)}.
\bibitem{Shen:2025mxe}
J.~F.~Shen, Y.~J.~Zhang and L.~Han,
\href{doi:10.1088/1674-1137/adf1ef}{Chin. Phys. C \textbf{49}, no.11, 113108  (2025)}.

\bibitem{Li:2025epjc}
R.~P.~Li, J.~W.~Lian and Y.~B. Liu,
 \href{doi:10.1140/epjc/s10052-025-15178-7}{Eur.\ Phys.\ J.\ C \textbf{85}, 1429 (2025)}.

\bibitem{Yue:2026tbc}
C.~X.~Yue, M.~S.~Y.~Wang, X.~Y.~Li and S.~Y.~Zhang,
[arXiv:2606.27878 [hep-ph]].
\bibitem{Sagheer:2026eql}
H.~Sagheer, I.~Ahmed and J.~Muhammad,
[arXiv:2606.28717 [hep-ph]].


\bibitem{Cui:2026wwo}
Y.~Cui, S.~Wang, Z.~H.~Yu and H.~H.~Zhang,
\href{doi:10.1103/p8l7-vv3p}{Phys. Rev. D \textbf{113} no.11, 115025 (2026) }.
[arXiv:2604.11232 [hep-ph]].


\bibitem{Liu:2026eyn}
Y.~B.~Liu and S.~Moretti,
[arXiv:2608.14195 [hep-ph]].

\bibitem{Thomas:1998wy}
S.~D.~Thomas and J.~D.~Wells,
\href{doi:10.1103/PhysRevLett.81.34}{Phys. Rev. Lett. \textbf{81}, 34-37 (1998)}.
[arXiv:hep-ph/9804359 [hep-ph]].


\bibitem{Cynolter:2008ea}
G.~Cynolter and E.~Lendvai,
\href{doi:10.1140/epjc/s10052-008-0771-7}{Eur. Phys. J. C \textbf{58}, 463-469 (2008)}.
[arXiv:0804.4080 [hep-ph]].


\bibitem{deBlas:2013gla}
J.~de Blas,
\href{doi:10.1051/epjconf/20136019008}{EPJ Web Conf. \textbf{60}, 19008 (2013)}.
[arXiv:1307.6173 [hep-ph]].

\bibitem{Adhikary:2024esf}
A.~Adhikary, M.~Olechowski, J.~Rosiek and M.~Ryczkowski,
\href{doi:10.1103/PhysRevD.110.075029}{Phys. Rev. D \textbf{110}, no.7, 075029  (2024)}.
[arXiv:2406.16050 [hep-ph]].

\bibitem{deBlas:2025pco}
J.~de Blas, C.~Giuliano, G.~Guedes, R.~S.~L{\'o}pez and J.~Santiago,
\href{doi:10.1007/JHEP05(2026)142}{JHEP \textbf{05}, 142 (2026)}.
[arXiv:2512.04148 [hep-ph]].

\bibitem{Daberstiel:2026wxe}
G.~Daberstiel, K.~M{\"o}hling, D.~St{\"o}ckinger and H.~St{\"o}ckinger-Kim,
\href{doi:10.1007/JHEP07(2026)259}{JHEP \textbf{07} (2026), 259}.
[arXiv:2603.21414 [hep-ph]].


\bibitem{Alwall:2014hca}
J. Alwall, R. Frederix, S. Frixione, V. Hirschi, F. Maltoni, O. Mattelaer, H.-S. Shao, T. Stelzer, P. Torrielli, and M. Zaro,
\href{doi:10.1007/JHEP07(2014)079}{JHEP \textbf{07}, 079 (2014)}.


\bibitem{Alloul:2013bka}
A.~Alloul, N.~D.~Christensen, C.~Degrande, C.~Duhr and B.~Fuks,
\href{doi:10.1016/j.cpc.2014.04.012}{Comput.\ Phys.\ Commun.\ \textbf{185}, 2250 (2014)}.
[arXiv:1310.1921 [hep-ph]].

\bibitem{Degrande:2011ua}
C.~Degrande, C.~Duhr, B.~Fuks, D.~Grellscheid, O.~Mattelaer and T.~Reiter,
\href{doi:10.1016/j.cpc.2012.01.022}{Comput.\ Phys.\ Commun.\ \textbf{183}, 1201 (2012)}.
[arXiv:1108.2040 [hep-ph]].

\bibitem{Bhattiprolu:VLL-UFO}
P.~N.~Bhattiprolu and S.~P.~Martin,
``VLL-UFOs: FeynRules and UFO model files for vector-like leptons,''
model package used in this work,
\href{https://github.com/prudhvibhattiprolu/VLL-UFOs}{github.com/prudhvibhattiprolu/VLL-UFOs}.




\bibitem{Sjostrand:2014zea}
T.~Sj\"ostrand, S. Ask, J. R. Christiansen {\it et al.},
\href{doi:10.1016/j.cpc.2015.01.024}{Comput. Phys. Commun. {\bf 191}, 159 (2015)}.


\bibitem{deFavereau:2013fsa}
J.~de Favereau \textit{et al.} (DELPHES 3 Collaboration),
\href{doi:10.1007/JHEP02(2014)057}{JHEP \textbf{02}, 057 (2014)}.
\bibitem{Leogrande:2019qbe}
E.~Leogrande, P.~Roloff, U.~Schnoor, and M.~Weber,
A DELPHES card for the CLIC detector,
\href{https://arxiv.org/abs/1909.12728}{arXiv:1909.12728}.
\bibitem{Boronat:2014hva}
M.~Boronat, J.~Fuster, I.~Garcia, E.~Ros and M.~Vos,
\href{doi:10.1016/j.physletb.2015.08.055}{Phys. Lett. B \textbf{750}, 95-99 (2015)}.
\bibitem{Boronat:2016tgd}
M.~Boronat, J.~Fuster, I.~Garcia, P.~Roloff, R.~Simoniello and M.~Vos,
\href{doi:10.1140/epjc/s10052-018-5594-6}{Eur.\ Phys.\ J.\ C \textbf{78}, 144 (2018)}.
\bibitem{Conte:2012fm}
E.~Conte, B.~Fuks, and G.~Serret,
\href{doi:10.1016/j.cpc.2012.09.009}
{Comput. Phys. Commun. \textbf{184}, 222 (2013)}.
[arXiv:1206.1599 [hep-ph]].
\bibitem{Cowan:2010js}
G.~Cowan, K.~Cranmer, E.~Gross and O.~Vitells,
\href{doi:10.1140/epjc/s10052-011-1554-0}{Eur.\ Phys.\ J.\ C \textbf{71}, 1554 (2011)}.
[Erratum: Eur.\ Phys.\ J.\ C \textbf{73}, 2501 (2013)].
[arXiv:1007.1727 [physics.data-an]].
\end{thebibliography}
\end{document}